\documentclass[a4paper,12pt]{article}
\usepackage[a4paper,text={16.8cm,22.4cm}]{geometry}
\usepackage{amsmath,amssymb,braket,bm}
\usepackage{cite}
\usepackage{graphicx}
\usepackage{hepunits}
\usepackage{color}
\usepackage{multirow}

\usepackage[utf8]{inputenc}

\allowdisplaybreaks
\DeclareMathOperator{\arccosh}{arccosh}
\DeclareMathOperator{\Li}{Li}
\DeclareMathOperator{\TO}{\mathbf{T}}
\DeclareMathOperator{\diag}{Diag}

\begin{document}

\begin{titlepage}

\begin{center}

\textbf{ \Large The next-to-next-to-leading order soft function for top quark pair production }

\vspace{7ex}

\textsc{ Guoxing Wang$^{a}$, Xiaofeng Xu$^{a}$, Li Lin Yang$^{a,b,c}$ and Hua Xing Zhu$^{d}$ }

\vspace{2ex}

\textsl{
${}^a$School of Physics and State Key Laboratory of Nuclear Physics and Technology,\\
Peking University, Beijing 100871, China
\\[0.3cm]
${}^b$Collaborative Innovation Center of Quantum Matter, Beijing, China
\\[0.3cm]
${}^c$Center for High Energy Physics, Peking University, Beijing 100871, China
\\[0.3cm]
${}^d$Zhejiang Institute of Modern Physics, Department of Physics, Zhejiang University, Hangzhou 310027, China
}
\end{center}

\vspace{4ex}

\begin{abstract}

We present the first calculation of the next-to-next-to-leading order threshold soft function for top quark pair production at hadron colliders, with full velocity dependence of the massive top quarks. Our results are fully analytic, and can be entirely written in terms of generalized polylogarithms. The scale-dependence of our result coincides with the well-known two-loop anomalous dimension matrix including the three-parton correlations, which at the two-loop order only appear when more than one massive partons are involved in the scattering process. In the boosted limit, our result exhibits the expected factorization property of mass logarithms, which leads to a consistent extraction of the soft fragmentation function. The next-to-next-to-leading order soft function obtained in this paper is an important ingredient for threshold resummation at the next-to-next-to-next-to-leading logarithmic accuracy.

\end{abstract}

\end{titlepage}

\section{Introduction}
\label{sec:intro}

Top quark pair production is one of the most important processes at the Large Hadron Collider (LHC). The total cross section at $\sqrt{s} = \unit{13}{\TeV}$ is about $\unit{800}{\picobarn}$. Both the ATLAS and the CMS experiments have collected nearly $\unit{100}{\invfb}$ of integrated luminosity at $\unit{13}{\TeV}$. This corresponds to 160 million $t\bar{t}$ events in total. With such a large event sample, top quark physics has become one of the precision frontier of particle physics. Many important measurements related to the top quark, e.g., inclusive and differential cross sections \cite{Aaboud:2017fha, Sirunyan:2017mzl}, top-quark mass and width \cite{Aaboud:2017ujq, Khachatryan:2014nda}, top-quark polarization \cite{Aaboud:2016hsq} and \textit{et al.}, can now be done at unprecedented precisions. The large production cross section also allows precision measurements of boosted top quark pairs, which is important for high-mass $t\bar{t}$ resonances search \cite{Chatrchyan:2012yca}, and for precision study of boosted top-quark jet \cite{Aad:2015hna}.

Currently, the best fixed-order calculations for top-quark pair production is at the next-to-next-to-leading order (NNLO) in QCD \cite{Baernreuther:2012ws, Czakon:2012zr, Czakon:2012pz, Czakon:2013goa, Czakon:2014xsa, Czakon:2015owf, Czakon:2016dgf} and the next-to-leading order (NLO) in the electroweak coupling \cite{Beenakker:1993yr, Bernreuther:2005is, Kuhn:2005it, Bernreuther:2006vg, Kuhn:2006vh, Hollik:2007sw, Bernreuther:2008md, Bernreuther:2010ny, Hollik:2011ps, Kuhn:2011ri, Bernreuther:2012sx, Kuhn:2013zoa, Campbell:2016dks, Pagani:2016caq, Denner:2016jyo}. Recently, these two corrections have been combined to give a more comprehensive description of $t\bar{t}$ production in \cite{Czakon:2017wor}. Despite the high precisions of these perturbative calculations, the complicated kinematics of $t\bar{t}$ production makes it necessary to consider even higher order corrections. This is particularly important since the high energy of the LHC has opened up the possibility to produce ``boosted'' top quark pairs, which means that the energies of the top quarks are much larger than their rest mass $m_t$. In \cite{Czakon:2016dgf}, it has been found that the NNLO QCD differential cross sections in the boosted regime are rather sensitive to the choice of factorization and renormalization scales. This scale dependence can be dramatically reduced by resumming certain towers of large logarithms to all orders in the strong coupling $\alpha_s$ \cite{Pecjak:2016nee, Czakon:2018nun}. These include not only the threshold logarithms which arise when the partonic center-of-mass energy approaches the $t\bar{t}$ invariant mass $M$, but also the mass logarithms of the form $\ln^n(m_t^2/M^2)$ which develop in the boosted region $M \gg m_t$.

The resummation of the threshold logarithms in $t\bar{t}$ production requires a couple of ingredients, such as the hard function and the soft function, as well as various anomalous dimensions. The NLO hard and soft functions have been computed in \cite{Ahrens:2010zv}, and the anomalous dimension matrices have been derived in \cite{Ferroglia:2009ep, Ferroglia:2009ii}. These enabled the resummation to be done at the next-to-next-to-leading logarithmic (NNLL) accuracy \cite{Ahrens:2010zv}. Given the NNLO accuracy achieved from fixed-order calculations, it is desirable to extend the threshold resummation to N$^3$LL. Such a calculation would improve the theoretical predictions over the whole phase space, all the way from low invariant mass to the boost regime. In order to achieve that, the NNLO hard and soft functions are necessary. The NNLO hard function can in principle be extracted from the virtual amplitude calculated in \cite{Baernreuther:2013caa}. Therefore, the NNLO soft function becomes a major bottleneck in pushing up the resummation accuracy of $t\bar{t}$ production, and is the subject of this article.

The soft functions describe the cross sections in the soft limit. The behavior of scattering amplitudes and cross sections in the soft limit is of high interest not only phenomenologically, but also theoretically. For example, the soft theorems in gauge theories and in gravitational theories \cite{Weinberg:1965nx, Gross:1968in, Jackiw:1968zza} are of fundamental importance in the understanding of their structures. In perturbative calculations in gauge theories, both the exchange of virtual soft particles and the emission of real soft ones can lead to infrared (IR) divergences. These must cancel against each other in order to arrive at meaningful predictions for physical observables. While such cancellations have been proven generically \cite{Kinoshita:1962ur, Lee:1964is}, the practical treatments of the IR divergences are highly non-trivial. Both the virtual and real contributions need to be calculated analytically in order to verify the precise cancellation. For the virtual amplitudes, when all external hard partons are massless, the soft singularities enjoy a dipole form up to two loops \cite{Aybat:2006wq}, thanks to the emerged scaling symmetry as the energy of hard partons become large \cite{Becher:2009cu, Gardi:2009qi}. Non-trivial corrections to the dipole form of soft singularities for massless scattering first appear at three loops, and have been computed recently in \cite{Almelid:2015jia}. The situation for massive amplitudes is much more complicated. Non-trivial correlations among three or four partons appear already at two loops \cite{Mitov:2009sv, Becher:2009kw, Ferroglia:2009ep, Ferroglia:2009ii, Chien:2011wz}. These virtual singularities must have the same structure as the real ones, and the soft functions provide a perfect place to investigate the latter. It is therefore interesting to calculate the massive soft functions through to NNLO, in order to study these multi-parton correlations from a different perspective.

The soft functions are defined as the vacuum expectation values of certain operators consisting of light-like and time-like soft Wilson lines. In simpler situations, they have been extensively studied in the literature. For processes involving two massless partons, such as the Drell-Yan process and Higgs production through gluon fusion, the soft functions have been calculated up to the N$^3$LO \cite{Li:2014afw}. For processes with 4 massless partons such as di-jet production and boosted heavy quark pair production, the NNLO soft function was obtained in \cite{Ferroglia:2012uy}. When massive partons are involved, the calculation gets much more complicated. The soft function for the $e^+ e^- \to t\bar{t}$ process has been calculated at the NNLO in \cite{vonManteuffel:2014mva}. Much less is known in the case of hadronic production of top quark pairs, for which only the NLO soft function is available \cite{Ahrens:2010zv}. Our result in this work therefore serves as the first example of an NNLO soft function for massive scattering with 4 external partons. 

This paper is organized as follows. In Section \ref{sec:form} we lay out the generic definition and renormalization of threshold soft functions. In Section \ref{sec:nlo} we provide the result of the NLO soft function to higher powers of the dimensional regulator $\epsilon$, which is a necessary ingredient in the renormalization of the soft function at NNLO. Section \ref{sec:nnlosoft} describes the main efforts of this work, namely the calculation of the NNLO bare soft function. We then perform its renormalization in Section \ref{sec:ren}, and discuss some cross-checks and the numerical impact of our new result. Finally, we conclude and discuss some future applications and extensions of our calculation in Section \ref{sec:con}.

\section{Formalism}
\label{sec:form}

We consider a generic scattering process involving energetic massless quarks, gluons and massive partons (such as top quarks or some new colored particles often present in models beyond the SM). The interactions of soft gluons with these energetic partons can be described by Wilson lines defined as
\begin{align}
\bm{S}_i(x) = \mathcal{P} \exp \! \left( ig_s \int_{-\infty}^0 ds \, v_i \cdot A^a(x+sv_i) \, \bm{T}_i^a \right) ,
\end{align}
where $\mathcal{P}$ denotes path ordering, $v_i$ is a 4-vector pointing to the direction of the momentum of the $i$-th parton, which satisfies $v_i^2 = 0$ for massless partons and $v_i^2>0$ for massive partons. Note that here we have taken all vectors $v_i$ to be incoming. The boldface $\bm{T}_i^a$ is the color generator associated with the $i$-th parton in the color-space formalism \cite{Catani:1996jh, Catani:1996vz}. It is evident that the Wilson lines are invariant under the rescaling $v_i \to \lambda v_i$ for any $\lambda > 0$, since this change can be compensated by a change of the integration variable $s \to s / \lambda$. We could employ this freedom to normalize the direction vectors of massive partons to $v_i^2 = 1$. This has the physical meaning that $v_i$ is the 4-velocity of the $i$-th parton: $v_i = \pm p_i / m_i$ where $m_i$ is the mass of the $i$-th parton. However, we'd like to keep this possibility open for the sake of generality.

Putting the Wilson lines together, the behavior of the $n$-parton scattering amplitude in the soft limit can be obtained via studying the vacuum matrix elements of the Wilson loop operator constructed out of the Wilson lines
\begin{align}
\bm{W}(x,\{v\}) \equiv \braket{0 | \bar{\TO} \! \left[ \bm{O}_s^\dagger(x) \right] \TO \! \left[ \bm{O}_s(0)\right] |0} \equiv \Braket{0 | \bar{\TO} \! \left[ \prod_{i=1}^n \bm{S}_i^\dagger(x) \right] \TO \! \left[ \prod_{i=1}^n \bm{S}_i(0) \right] |0} \, ,
\end{align}
where $\{v\}$ denotes the collection of the directional vectors $v_i$, $x$ is a time-like vector, and $\TO$ denotes time-ordering. It is well-known that the vacuum matrix elements of the Wilson loop operator, when calculated in perturbation theory, contain ultraviolet (UV) divergences which need to be renormalized \cite{Korchemsky:1987wg, Korchemskaya:1992je}. The renormalization properties of the Wilson loops can be used to study the infrared singularities of scattering amplitudes, as was illustrated in \cite{Becher:2009cu, Becher:2009qa, Becher:2009kw, Ferroglia:2009ep, Ferroglia:2009ii}.

The Wilson loop operator is also an essential ingredient in the factorization of scattering cross sections in the soft limit. We consider scattering processes at hadron-hadron colliders with no final state massless partons at the leading order. These include, for example, top quark pair production (possibly associated with other colorless particles such as the Higgs boson and electroweak gauge bosons), production of 4 top quarks, squark and gluino productions in supersymmetric models, as well as productions of top partners in many new physics models. At higher orders in the strong coupling constant, there will be additional emissions of gluons and quarks in the final state. We are interested in the case where these additional emissions are all soft, i.e., with energies much smaller than the typical momentum transfer of the hard-scattering process. Note that the precise meaning of ``soft'' depends on the reference frame, which leads to different forms of the factorization formula, such as the ``pair-invariant-mass'' (PIM) kinematics and the ``single-particle-inclusive'' (1PI) kinematics in top quark pair production discussed, e.g., in \cite{Ahrens:2010zv, Kidonakis:2010dk, Ahrens:2011mw}. While the formalism can be applied to any reference frame, in the following, we will work in the center-of-mass frame of the two incoming partons, which are not only good for demonstration purposes, and are also adopted in many existing calculations. For example, this corresponds to the PIM kinematics in \cite{Kidonakis:1996aq, Kidonakis:1997gm, Ahrens:2010zv} for $t\bar{t}$ production, in \cite{Broggio:2013uba, Broggio:2013cia} for stop pair production, and in \cite{Broggio:2015lya, Broggio:2016lfj} for $t\bar{t}H$ production. Schematically, we are considering partonic processes of the form
\begin{align}
  h_1(p_1) + h_2(p_2) \to h_3(p_3) + \cdots + h_n(p_n) + X + X_s(p_s) \, ,
\end{align}
where $h_1$ and $h_2$ are two incoming massless partons, $h_I$ ($I=3,\ldots,n$) are outgoing massive partons, $X$ denotes colorless particles such as the Higgs boson and electroweak gauge bosons, and $X_s$ represents additional soft radiations which we want to describe. Here, the momenta $p_I$ ($I=3,\ldots,n$) are chosen to be outgoing, but we still keep $v_I$ to be incoming for convenience.

In the center-of-mass frame of the two incoming partons, the emissions of additional soft partons are described by the so-called soft function $\bm{S}$, which is simply the momentum-space version of the vacuum matrix element of the Wilson loop operator:
\begin{align}
\bm{S}(\omega,\{v\}) &\equiv \frac{1}{\sqrt{d_1d_2}} \int \frac{dx_0}{4\pi} \, e^{i\omega x_0/2} \, \Big[ \bm{W}(x,\{v\}) \Big]_{x_\mu=(x_0,0,0,0)} \nonumber
\\
\label{eq:Smom}
&= \frac{1}{\sqrt{d_1d_2}} \sum_{X_s} \braket{0 | \bm{O}_s^\dagger(0) | X_s} \braket{X_s | \bm{O}(0) | 0} \delta ( \omega - v_0 \cdot p_{X_s} ) \, ,
\end{align}
where the reference vector $v_0=(2,0,0,0)$, and $\omega$ represents (2 times) the energy of the additional soft partons. We have included a normalization factor such that the definition of the soft function coincides with that in \cite{Ahrens:2010zv}. Here $d_1$ and $d_2$ are the dimensions of the $SU(N)_{\text{color}}$ representations to which the partons $h_1$ and $h_2$ belong. For later convenience, it is useful to perform a Mellin or Laplace transform into the moment space
\begin{align}
\label{eq:laplace}
\tilde{\bm{s}}(\Lambda,\{v\}) &= \int_0^\infty d\omega \, \exp \! \left( -\frac{\omega}{\Lambda e^{\gamma_E}} \right) \bm{S}(\omega,\{v\}) =  \frac{1}{\sqrt{d_1d_2}} \, \bm{W} \! \left( x_0 = \frac{-2i}{\Lambda e^{\gamma_E}} , \{v\} \right) ,
\end{align}
where $\Lambda$ is a soft momentum scale in the moment space. As discussed above, the bare soft function contains UV divergences which can be regularized in dimensional regularization with $d=4-2\epsilon$. These UV divergences are removed by the operator renormalization
\begin{align}
\tilde{\bm{s}}(L,\{v\},\mu) = \bm{Z}_s^\dagger(L,\{v\},\mu) \, \tilde{\bm{s}}_{\text{bare}}(\Lambda,\{v\}) \, \bm{Z}_s(L,\{v\},\mu) \, ,
\end{align}
where $\mu$ is the renormalization scale and $L\equiv\ln(\Lambda^2/\mu^2)$. As indicated in the above formula, both the renormalized soft function $\tilde{\bm{s}}(L,\{v\},\mu)$ and the renormalization factor $\bm{Z}_s(L,\{v\},\mu)$ are $\mu$-dependent and satisfy renormalization group equations (RGEs)
\begin{align}
\label{eq:zrge}
\frac{d}{d\mu} \bm{Z}_s(L,\{v\},\mu) &= - \bm{Z}_s(L,\{v\},\mu) \, \bm{\Gamma}_s(L,\{v\},\mu) \, ,
\\
\label{eq:srge}
\frac{d}{d\mu} \tilde{\bm{s}}(L,\{v\},\mu) &= - \bm{\Gamma}_s^\dagger(L,\{v\},\mu) \, \tilde{\bm{s}}(L,\{v\},\mu) - \tilde{\bm{s}}(L,\{v\},\mu) \, \bm{\Gamma}_s(L,\{v\},\mu) \, .
\end{align}
The generic form of the soft anomalous dimension matrix $\bm{\Gamma}_s$ up to the two-loop order can be written as \cite{Becher:2009kw}
\begin{align}
\label{eq:gammaS}
\bm{\Gamma}_s(L,\{v\},\mu) &= \frac{\bm{T}_1^2 + \bm{T}_2^2}{2} \, \gamma_{\text{cusp}}(\alpha_s) \, L_s - \sum_{(I,J)} \frac{\bm{T}_I \cdot \bm{T}_J}{2} \, \gamma_{\text{cusp}}(\beta_{IJ},\alpha_s) + \sum_i \gamma^i_s(\alpha_s) + \sum_I \gamma^I(\alpha_s) \nonumber
\\
&\, + \sum_{I} \left[ \bm{T}_I \cdot \bm{T}_1 \, \gamma_{\text{cusp}}(\alpha_s) \, \ln\frac{v_1 \cdot v_2 \, \sqrt{v_I^2}}{v_0 \cdot v_2 \; w_{I1}} + \bm{T}_I \cdot \bm{T}_2 \, \gamma_{\text{cusp}}(\alpha_s) \, \ln\frac{v_1 \cdot v_2 \, \sqrt{v_I^2}}{v_0 \cdot v_1 \; w_{I2}} \right] \nonumber
\\
&\, + \sum_{(I,J,K)} if^{abc} \, \bm{T}_I^a \, \bm{T}_J^b \, \bm{T}_K^c \, F_1(\beta_{IJ},\beta_{JK},\beta_{KI}) \nonumber
\\
&\, + \sum_{(I,J)} \sum_k if^{abc} \, \bm{T}_I^a \, \bm{T}_J^b \, \bm{T}_k^c \, f_2 \! \left( \beta_{IJ}, \ln\frac{w_{Jk} \, \sqrt{v_I^2}}{w_{Ik} \, \sqrt{v_J^2}} \right) + \mathcal{O}(\alpha_s^3) \, ,
\end{align}
where the lower-case indices ($i$, $j$, $k$) run over massless partons 1 and 2, while the capital indices ($I$, $J$, $K$) run over massive partons. We have introduced the abbreviations $L_s \equiv L - i\pi$ and $w_{Ij} \equiv v_I \cdot v_j + i0$. The notation $(I,J,\ldots)$ denotes unordered tuples of distinct parton indices. The functions $\gamma_{\text{cusp}}(\alpha_s)$ and $\gamma_{\text{cusp}}(\beta_{IJ},\alpha_s)$ are the famous cusp anomalous dimensions for light-like Wilson lines and time-like Wilson lines, respectively \cite{Korchemsky:1987wg, Korchemsky:1991zp, Kidonakis:2009ev}
\begin{align}
\gamma_{\text{cusp}}(\alpha_s) &= \frac{\alpha_s}{\pi} + \left( \frac{\alpha_s}{4\pi} \right)^2 \left[ \left( \frac{268}{9} - \frac{4\pi^2}{3} \right) C_A - \frac{80}{9} \, T_F N_l \right] + \mathcal{O}(\alpha_s^3) \, ,
\\
\gamma_{\text{cusp}}(b,\alpha_s) &= \gamma_{\text{cusp}}(\alpha_s) \, b \coth(b) \nonumber
\\
&\hspace{-3em} + \frac{C_A}{2} \left( \frac{\alpha_s}{\pi} \right)^2 \left\{ \frac{\pi^2}{6} + \zeta_3 + b^2 + \coth^2(b) \left[ \Li_3(e^{-2b}) + b \Li_2(e^{-2b}) - \zeta_3 + \frac{\pi^2}{6} b + \frac{b^3}{3} \right] \right.
\\
&\hspace{1em} \left. + \coth(b) \left[ \Li_2(e^{-2b}) - 2 b \ln(1-e^{-2b}) - \frac{\pi^2}{6} (1+b) - b^2 - \frac{b^3}{3} \right] \right\} + \mathcal{O}(\alpha_s^3) \, . \nonumber
\end{align}
with the cusp angle given by
\begin{align}
\beta_{IJ} = \arccosh \! \left( - \frac{v_I \cdot v_J}{\sqrt{v_I^2 \, v_J^2}} - i0 \right) \, .
\end{align}
The single parton soft anomalous dimensions $\gamma_s^i(\alpha_s)$ and $\gamma^I(\alpha_s)$ are
\begin{align}
\gamma_s^i(\alpha_s) &= \left( \frac{\alpha_s}{4\pi} \right)^2 C_i \left[  \left( -\frac{404}{27} + \frac{11\pi^2}{18} + 14\zeta_3 \right) C_A + \left( \frac{112}{27} - \frac{2\pi^2}{9} \right) T_F N_l \right] + \mathcal{O}(\alpha_s^3) \, ,
\\
\gamma^I(\alpha_s) &= -C_I \frac{\alpha_s}{2\pi} + \left( \frac{\alpha_s}{4\pi} \right)^2 C_I \left[ \left( -\frac{98}{9} + \frac{2\pi^2}{3} - 4\zeta_3 \right) C_A + \frac{40}{9} \, T_F N_l \right] + \mathcal{O}(\alpha_s^3) \, ,
\end{align}
where $C_{i(I)}=C_F$ for the fundamental representation, and $C_{i(I)}=C_A$ for the adjoint representation of the gauge group. The three-parton correlation functions $F_1$ and $f_2$ were calculated in \cite{Ferroglia:2009ep, Ferroglia:2009ii}. The function $F_1$ describes correlations among three massive partons, and can be written as
\begin{align}
F_1(\beta_{IJ},\beta_{JK},\beta_{JI}) &= \left( \frac{\alpha_s}{4\pi} \right)^2 \, \frac{4}{3} \sum_{L,M,N} \epsilon_{LMN} \, g(\beta_{LM}) \, \beta_{NL} \coth(\beta_{NL}) + \mathcal{O}(\alpha_s^3) \, ,
\end{align}
where the indices ($L$,$M$,$N$) run over ($I$,$J$,$K$) with $\epsilon_{LMN}=1$ if ($L$,$M$,$N$) is an even permutation of ($I$,$J$,$K$), and
\begin{align}
g(b) = \coth(b) \left[ b^2 + 2 b \ln(1-e^{-2b}) - \Li_2(e^{-2b}) + \frac{\pi^2}{6} \right] - b^2 - \frac{\pi^2}{6} \, .
\end{align}
The function $f_2$ describes correlations among two massive partons and one massless parton, and is given by
\begin{align}
f_2 \! \left( \beta_{IJ}, \ln\frac{w_{Jk} \, \sqrt{v_I^2}}{w_{Ik} \, \sqrt{v_J^2}} \right) &= - \left( \frac{\alpha_s}{4\pi} \right)^2 4 g(\beta_{IJ}) \times \ln\frac{w_{Jk} \, \sqrt{v_I^2}}{w_{Ik} \, \sqrt{v_J^2}} + \mathcal{O}(\alpha_s^3) \, ,
\end{align}

Given the anomalous dimension matrix $\bm{\Gamma}_s$, one can solve the RGE (\ref{eq:zrge}) to obtain the renormalization factor $\bm{Z}_s$. To this end, it is useful to decompose $\bm{\Gamma}_s$ in (\ref{eq:gammaS}) into the form
\begin{align}
\bm{\Gamma}_s(L,\{v\},\mu) \equiv \frac{\alpha_s}{4\pi} \left( A_0 L_s + \bm{\Gamma}_0 \right) + \left( \frac{\alpha_s}{4\pi} \right)^2 \left( A_1 L_s + \bm{\Gamma}_1 \right) ,
\end{align}
and then
\begin{align}
\label{eq:Zs}
\ln \bm{Z}_s(L,\{v\},\mu) &= \frac{\alpha_s}{4\pi} \left( -\frac{A_0}{2\epsilon^2} + \frac{A_0 L_s + \bm{\Gamma}_0}{2\epsilon} \right) \nonumber
\\
&\, + \left( \frac{\alpha_s}{4\pi} \right)^2 \left[ \frac{3A_0\beta_0}{8\epsilon^3} + \frac{-A_1 - 2\beta_0 (A_0 L_s + \bm{\Gamma}_0)}{8\epsilon^2} + \frac{A_1 L_s + \bm{\Gamma}_1}{4\epsilon} \right] + \mathcal{O}(\alpha_s^3) \, ,
\end{align}
where $\beta_0 = (11 C_A - 4 T_F N_l) / 3$.

\section{The soft function for $t\bar{t}$ production and the NLO result to arbitrary orders in $\epsilon$}
\label{sec:nlo}

While the formalism introduced in the last section is very generic and applies to a lot of processes, the actual calculation of the soft function could get very complicated when the number of independent scalar products $v_i \cdot v_j$ becomes large. In this paper, we begin with the special case of $t\bar{t}$ production, where the partonic processes can be described as
\begin{align}
  h_1(p_1) + h_2(p_2) \to t(p_3) + \bar{t}(p_4) + X_s(p_s) \, ,
\end{align}
where $p_3^2 = p_4^2 = m_t^2$ with $m_t$ the mass of the top quark. In the soft limit $p_s \to 0$, there are 3 independent Lorentz invariant kinematic variables, which can be chosen as $m_t$ and
\begin{align}
M^2 \equiv (p_1+p_2)^2 \, , \quad t_1 \equiv (p_1-p_3)^2-m_t^2 \, .
\end{align}
It is convenient to introduce dimensionless quantities $\beta$ and $y = \cos\theta$, defined as
\begin{align}
\beta = \sqrt{1-\frac{4m_t^2}{M^2}} \, , \quad t_1 = -\frac{M^2}{2} (1-\beta y) \, ,
\end{align}
where $\beta$ and $\theta$ have the physical meanings of the 3-velocity and the scattering angle of the top quark in the partonic center-of-mass frame. The soft function depends on $\beta$ and $y$ through the following combinations of the directional vectors $v_i$:
\begin{align}
\frac{v_3 \cdot v_1 \, v_2 \cdot v_0}{\sqrt{v_3^2} \, v_1 \cdot v_2} &= \frac{v_4 \cdot v_2 \, v_1 \cdot v_0}{\sqrt{v_4^2} \, v_1 \cdot v_2} = - \frac{1-\beta y}{\sqrt{1-\beta^2}} \, , \quad \frac{v_3 \cdot v_4}{\sqrt{v_3^2 \, v_4^2}} = \frac{1+\beta^2}{1-\beta^2} \, , \nonumber
\\
\frac{v_3 \cdot v_2 \, v_1 \cdot v_0}{\sqrt{v_3^2} \, v_1 \cdot v_2} &= \frac{v_4 \cdot v_1 \, v_2 \cdot v_0}{\sqrt{v_4^2} \, v_1 \cdot v_2} = -\frac{1+\beta y}{\sqrt{1-\beta^2}} \, ,
\end{align}
and we also have
\begin{align}
\beta_{34} \equiv \arccosh \! \left( -\frac{v_3 \cdot v_4}{\sqrt{v_3^2 \, v_4^2}} - i0 \right) = \ln\frac{1+\beta}{1-\beta} - i\pi \, .
\end{align}

To calculate the bare soft function, it is convenient to start from the momentum-space version $\bm{S}_{\text{bare}}(\omega,\beta,y)$ introduced in Eq.~(\ref{eq:Smom}). Here we have expressed the dependence on the directional vectors $v_i$ through the quantities $\beta$ and $y$. The perturbative expansion of the momentum-space soft function can be written as
\begin{align}
\bm{S}_{\text{bare}}(\omega,\beta,y) = \delta(\omega) \, \frac{\bm{1}}{\sqrt{d_1d_2}} + \frac{\alpha_s}{4\pi} \, \bm{S}_{\text{bare}}^{(1)}(\omega,\beta,y) + \left( \frac{\alpha_s}{4\pi} \right)^2 \bm{S}_{\text{bare}}^{(2)}(\omega,\beta,y) + \cdots \, ,
\end{align}
where $\bm{1}$ denotes the identity operator in color space, and the NLO soft function $\bm{S}_{\text{bare}}^{(1)}(\omega,\beta,y)$ was already calculated in \cite{Ahrens:2010zv}. In fact, since the NLO soft function involves 2-parton correlations at most, the same calculation can be applied to scattering processes with more than two colored particles in the final state. This has been done, e.g., for the case of $t\bar{t}H$ production in \cite{Broggio:2015lya, Broggio:2016lfj}, and can be extended to more complicated processes such the simultaneous production of two $t\bar{t}$ pairs. In order to calculate the NNLO soft function, however, we will need the NLO one to higher orders in the dimensional regulator $\epsilon$, which will produce a finite contribution to the renormalized NNLO soft function. We will describe such a calculation in the following, and the calculation of the NNLO bare soft function will be discussed in Section~\ref{sec:nnlosoft}.

In \cite{Ahrens:2010zv}, the NLO soft function was calculated by brute-force evaluation of the relevant phase-space integrals. While it is possible to continue using such a method to obtain the higher order terms in $\epsilon$, it is useful to employ a more systematic approach which can be extended to the NNLO calculation.

The definition (\ref{eq:Smom}) of the soft function involves a summation over soft final states $X_s$. It is easy to see that when $\ket{X_s}$ is the vacuum state, i.e., when there is no extra soft emission, the matrix elements involve scaleless integrals in dimensional regularization, which are defined to be zero. Therefore, at the NLO, the only contribution is given by
\begin{align}
\frac{\alpha_s}{4\pi} \, \bm{S}_{\text{bare}}^{(1)}(\omega,\beta,y) = \frac{1}{\sqrt{d_1d_2}} \int \frac{d^dk}{(2\pi)^{d-1}} \, \delta^+(k^2) \braket{0 | \bm{O}_s^\dagger(0) | g(k)} \braket{g(k) | \bm{O}(0) | 0} \delta ( \omega - v_0 \cdot k ) \, ,
\end{align}
where a summation over the helicity and the color of the gluon is understood. We use \texttt{QGRAF} \cite{Nogueira:1991ex} to generate the squared amplitudes in the above formula, and use \texttt{FORM} \cite{Vermaseren:2000nd} to manipulate the resulting expression. The phase-space integrals appearing in the result have the general form
\begin{align}
\label{eq:Iij}
I_{ij}(\epsilon,\omega,\beta,y) = \int [dk] \, \frac{v_i \cdot v_j}{v_i \cdot k \, v_j \cdot k} \, \delta( \omega - v_0 \cdot k ) \, ,
\end{align}
where $[dk] \equiv d^dk \, \delta^+(k^2)$. From symmetry considerations, it is obvious that we only need to calculate $I_{12}$, $I_{13}$, $I_{33}$ and $I_{34}$. At this point, we note that while the soft function itself does not depend on the normalizations of the directional vectors $v_i$, it is convenient to fix them in practical calculations. Therefore we will choose the normalizations $v_1 \cdot v_0 = v_2 \cdot v_0 = v_1 \cdot v_2 = 2$, and $v_3^2=v_4^2 = 1-\beta^2$ in the following. Note that the normalizations of $v_3$ and $v_4$ are unconventional. In the center-of-mass frame of the incoming partons, these vectors are then parameterized by
\begin{gather}
v_1 = (1,0,0,1) \, , \quad v_3 = - (1,0,\beta\sin\theta,\beta y) \, ,\nonumber
\\
v_2 = (1,0,0,-1) \, , \quad v_4 = - (1,0,-\beta\sin\theta,-\beta y) \, .
\end{gather}

The phase-space integrals appearing in the result are not independent, and we employ the integration-by-parts (IBP) \cite{Chetyrkin:1981qh, Tkachov:1981wb} method to find relations among them. To this end it is necessary to use the relation
\begin{align}
\label{eq:cutkosky1}
\delta^+(k^2) \equiv \delta(k^2) \, \theta(k^0) = \frac{1}{2\pi i} \left( \frac{1}{k^2 + i0} - \frac{1}{k^2 - i0} \right) ,
\end{align}
which is known as the reverse unitarity method \cite{Anastasiou:2002yz}, to express the phase-space integrals in terms of loop integrals. The $\delta$-function in Eq.~(\ref{eq:Iij}) can be similarly written as
\begin{align}
\label{eq:cutkosky2}
\delta(\omega - v_0 \cdot k) = \frac{1}{2\pi i} \left( \frac{1}{\omega - v_0 \cdot k + i0} - \frac{1}{\omega - v_0 \cdot k - i0} \right)
\end{align}
These integrals are then feed into the program packages \texttt{Reduze2} \cite{vonManteuffel:2012np} and \texttt{FIRE5} \cite{Smirnov:2014hma}, which use the IBP relations to reduce the relevant loop integrals to a number of master integrals. After the IBP reduction, one can recover the phase-space integrals by reversing the relations (\ref{eq:cutkosky1}) and (\ref{eq:cutkosky2}). In the NLO case, the master integrals can be chosen as $F_{0,0}$, $F_{0,1}$ and $F_{1,1}$, where $F_{a_1,a_2}$ is defined as
\begin{align}
F_{a_1,a_2} \equiv \int [dk] \, \delta ( \omega - v_0 \cdot k ) \, \frac{1}{(v_1 \cdot k)^{a_1} \, (-v_3 \cdot k)^{a_2}} \, .
\end{align}

In order to calculate the master integrals to arbitrary orders in the dimensional regulator $\epsilon$, we employ the method of differential equations \cite{Kotikov:1990kg, Gehrmann:1999as}. Taking the partial derivative of $F_{a_1,a_2}$ with respect to $\beta$ will lead to integrals with $a_2$ index shifted. However, all these integrals can be expressed as linear combinations of the master integrals. We collect the master integrals into a vector with 3 components
\begin{align}
\vec{f}(\epsilon,\beta,y,\omega) \equiv ( F_{0,0}, F_{0,1}, F_{1,1} )^{\mathsf{T}} \, .
\end{align}
The partial derivative of $\vec{f}$ with respect to $\beta$ then has the form
\begin{align}
\partial_\beta \vec{f}(\epsilon, \beta, y, \omega) = \hat{A}(\epsilon, \beta, y, \omega) \, \vec{f}(\epsilon, \beta, y, \omega) \, ,
\end{align}
where $\hat{A}$ is a $3 \times 3$ matrix. The matrix $\hat{A}$ is a rather complicated function of $\epsilon$, $\beta$, $y$ and $\omega$, which makes the differential equation not so straightforward to solve. It is possible to simplify the above equation by a linear transformation $\vec{g}(\epsilon,\beta,y) = \hat{T}(\epsilon, \beta, y, \omega) \, \vec{f}(\epsilon, \beta, y, \omega)$, where the matrix $\hat{T}$ is given by
\begin{align}
  \hat{T}(\epsilon, \beta, y, \omega) = \frac{2 \, \Gamma(1-2\epsilon)}{\pi^{1-\epsilon} \, \omega^{1-2\epsilon} \, \Gamma(1-\epsilon)}
  \begin{pmatrix}
    1-2\epsilon & 0 & 0
    \\
    0 & \epsilon \, \omega \, \beta & 0
    \\
    0 & 0 & \epsilon \, \omega^2 \, (1 - \beta y)
  \end{pmatrix}
  \, .
\end{align}
The new vector $\vec{g}$ (so-called ``canonical basis'') satisfies a simpler differential equation \cite{Henn:2013pwa}
\begin{align}
\label{eq:nloDEQ}
\partial_\beta \vec{g}(\epsilon,\beta,y) = \epsilon \, \hat{B}(\beta,y) \, \vec{g}(\epsilon,\beta,y) \, .
\end{align}
Note that the matrix $\hat{B}$ does not depend on $\epsilon$ and $\omega$ anymore, and is given by
\begin{align}
\hat{B}(\beta,y) = -\frac{\hat{a}}{\beta-1} + \frac{\hat{b}}{\beta} + \frac{\hat{c}}{\beta+1} - \frac{\hat{d}}{\beta-1/y} \, ,
\end{align}
where
\begin{align}
  \hat{a} =
  \begin{pmatrix}
    0 & 0 & 0
    \\
    1 & 1 & 0
    \\
    2 & 2 & 0
  \end{pmatrix}
  , \quad
  \hat{b} =
  \begin{pmatrix}
    0 & 0 & 0
    \\
    0 & 2 & 0
    \\
    4 & 0 & 2
  \end{pmatrix}
  , \quad
  \hat{c} =
  \begin{pmatrix}
    0 & 0 & 0
    \\
    1 & -1 & 0
    \\
    -2 & 2 & 0
  \end{pmatrix}
  , \quad
  \hat{d} =
  \begin{pmatrix}
    0 & 0 & 0
    \\
    0 & 0 & 0
    \\
    0 & 0 & 2
  \end{pmatrix}
  .
\end{align}
The vector $\vec{g}$ has no singularity in $\epsilon$ and can be Taylor-expanded in the form
\begin{align}
\vec{g}(\epsilon,\beta,y) = \sum_{n=0}^{\infty} \vec{g}^{(n)}(\beta,y) \, \epsilon^n \, .
\end{align}
The differential equation (\ref{eq:nloDEQ}) can then be solved order by order in $\epsilon$:
\begin{align}
\label{eq:nloDEQ2}
\vec{g}^{(n+1)}(\beta,y) = \int_{\beta_0}^\beta d\beta' \, \hat{B}(\beta',y) \, \vec{g}^{(n)}(\beta',y) + \vec g^{(n+1)}(\beta_0,y) \, ,
\end{align}
where the boundary conditions $\vec{g}^{(n)}(\beta_0,y)$ at some boundary point $\beta_0$ need to be obtained through other methods, which will be discussed later. Such iterated integrals give rise to so-called generalized polylogarithms (GPLs) \cite{Goncharov:1998kja}, which are defined by
\begin{align}
G(a_1,\ldots,a_n;\beta) \equiv \int_0^\beta \frac{d\beta'}{\beta'-a_1} G(a_2,\ldots,a_n;\beta') \, ,
\end{align}
with $G(;\beta) \equiv 1$. The special case where all the $a_i$'s are zero is defined as
\begin{align}
G(0,\ldots,0;\beta) \equiv \frac{1}{n!} \log^n\beta \, .
\end{align}
The GPLs have many good mathematical properties (for a review, see e.g. \cite{Duhr:2014woa}), and can be straightforwardly evaluated by program packages such as \texttt{GiNaC} \cite{Bauer:2000cp}. They therefore form a wonderful basis for expressing our results.

In order to solve the differential equations (\ref{eq:nloDEQ}), we also need the explicit expression of $\vec{g}(\epsilon,\beta,y)$ at the boundary point $\beta_0$, serving as the boundary condition. In our case, it is convenient to choose the point $\beta = 0$ as the boundary. The calculation of the boundary condition can be simplified by observing that the matrix $\hat{B}$ contains a singular term proportional to $1/\beta$. It is clear that $\partial_\beta F_{a_1,a_2}$ can produce a $1/\beta$ coefficient only if $F_{a_1,a_2}$ itself develops a power-like or logarithmic divergence when $\beta \to 0$. One can easily check that all the master integrals in $\vec{f}$ are regular in the limit $\beta \to 0$. The same applies to the components of $\vec{g}$ since the transformation matrix $\hat{T}$ is also regular. It follows that
\begin{align}
0 = \lim_{\beta \to 0} \beta \, \partial_\beta \vec{g}(\epsilon,\beta,y) = \lim_{\beta \to 0} \beta \, \epsilon \, \hat{B}(\beta,y) \, \vec{g}(\epsilon,\beta,y) \, ,
\end{align}
which leads to the conditions
\begin{align}
\label{eq:boundaryrelation}
g_3(\beta=0) = -2g_1(\beta=0) \, , \quad g_2(\beta=0) = 0 \, ,
\end{align}
with $g_i$ being the $i$-th component of $\vec{g}$. We now only need to directly evaluate the component $g_1$ (which actually doesn't depend on $\beta$) at the boundary $\beta = 0$, which is very simple:
\begin{align}
g_1(\epsilon,0,y) = \frac{2 \, \Gamma(2-2\epsilon)}{\pi^{1-\epsilon} \, \omega^{1-2\epsilon} \, \Gamma(1-\epsilon)} \int [dk] \, \delta(\omega-v_0\cdot k) = 1 \, .
\end{align}

We now have everything we need to express the NLO soft function as an abstract matrix in color space, in terms of the inner products of color generators $\bm{T}_i \cdot \bm{T}_j \equiv \bm{T}^a_i \, \bm{T}^a_j$. This abstract form is generic and especially useful if we want to apply our result to more complicated processes. However, for practical computations of $t\bar{t}$ cross sections, it is convenient to choose a color basis and express the soft function as a $2 \times 2$ matrix in the quark-anti-quark annihilation channel, and a $3 \times 3$ matrix in the gluon fusion channel. Such matrix elements are defined as
\begin{align}
\bm{S}_{IJ}(\omega,\beta,y) \equiv \braket{c_I | \bm{S}(\omega,\beta,y) | c_J} \, ,
\end{align}
where $\{\ket{c_I}\}$ is an orthogonal color basis. In accordance with \cite{Ahrens:2010zv}, we choose the singlet-octet basis with
\begin{gather}
\left(c_1^{q\bar{q}}\right)_{\{a\}} = \delta_{a_1a_2} \delta_{a_3a_4} \, , \quad \left(c_2^{q\bar{q}}\right)_{\{a\}} = t^c_{a_2a_1} t^c_{a_3a_4} \, , \nonumber
\\
\label{eq:colorbasis}
\left(c_1^{gg}\right)_{\{a\}} = \delta^{a_1a_2} \delta_{a_3a_4} \, , \quad \left(c_2^{gg}\right)_{\{a\}} = if^{a_1a_2c} \, t^c_{a_3a_4} \, , \quad \left(c_3^{gg}\right)_{\{a\}} = d^{a_1a_2c} \, t^c_{a_3a_4} \, ,
\end{gather}
where $a_i$ is the color index of the $i$-th parton. We have compared the resulting NLO matrices with those (up to order $\epsilon^0$) in \cite{Ahrens:2010zv} and find complete agreement.

\section{The NNLO bare soft function}
\label{sec:nnlosoft}

We now turn to the calculation of the NNLO bare soft function, which is the main new result of our paper. The contributions to the bare NNLO soft function consist of two parts: the virtual-real diagrams and the double-real diagrams. The two-loop virtual diagrams leading to vanishing integrals in dimensional regularization and we do not need to consider them.

\subsection{Double-real contributions}

We first present the calculation of the double-real contributions. As in the NLO calculation, we generate relevant Feynman diagrams and amplitudes using \texttt{QGRAF} \cite{Nogueira:1991ex}. The phase-space integrals in the double-real contribution have the generic form
\begin{align}
\int [dk_1] \, [dk_2] \, \delta \big( \omega - v_0 \cdot (k_1+k_2) \big) \, \mathcal{F}(\{v\},k_1,k_2) \, ,
\end{align}
where $\mathcal{F}$ denotes the integrand consisting of scalar products among the directional vectors $v_i$ and the two momenta $k_1$ and $k_2$. We generically call these scalar products ``propagators''. There exist many different propagators in our squared amplitudes. However, only a subset of them appears in any individual integral. It is therefore useful to classify all the integrals into a couple of ``integral families'', each defined by a particular set of propagators. For this purpose, we first classify the relevant Feynman diagrams into three categories according to the number of independent Wilson lines involved: 1) those involving one or two Wilson lines; 2) those involving three Wilson lines; and 3) those involving all four Wilson lines. We discuss the calculation of the first two categories in the following. The diagrams involving all four Wilson lines can be trivially expressed as a convolution of two NLO integrals, and we do not bother to discuss them here.

\subsubsection{One-  or two-Wilson-line diagrams}

\begin{figure}[t!]
\begin{center}
\includegraphics[width=0.6\textwidth]{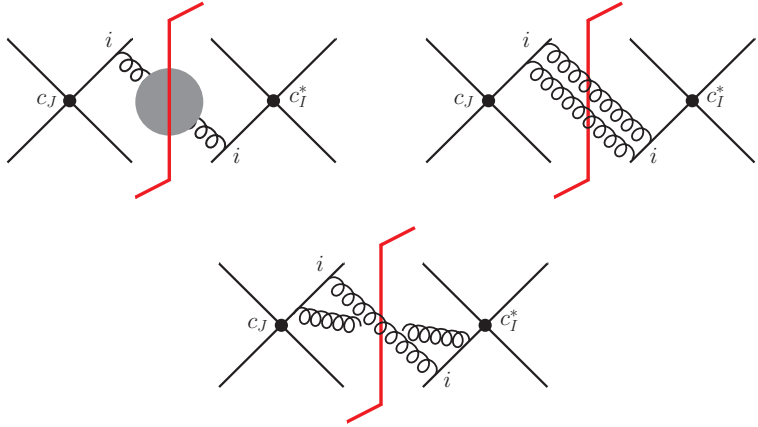}
\end{center}
\vspace{-2ex}
\caption{One-Wilson-line double-real diagrams contributing to the NNLO soft function.}
\label{fig:NNLORRoneWilsonLine}
\end{figure}

\begin{figure}[t!]
\begin{center}
\includegraphics[width=0.6\textwidth]{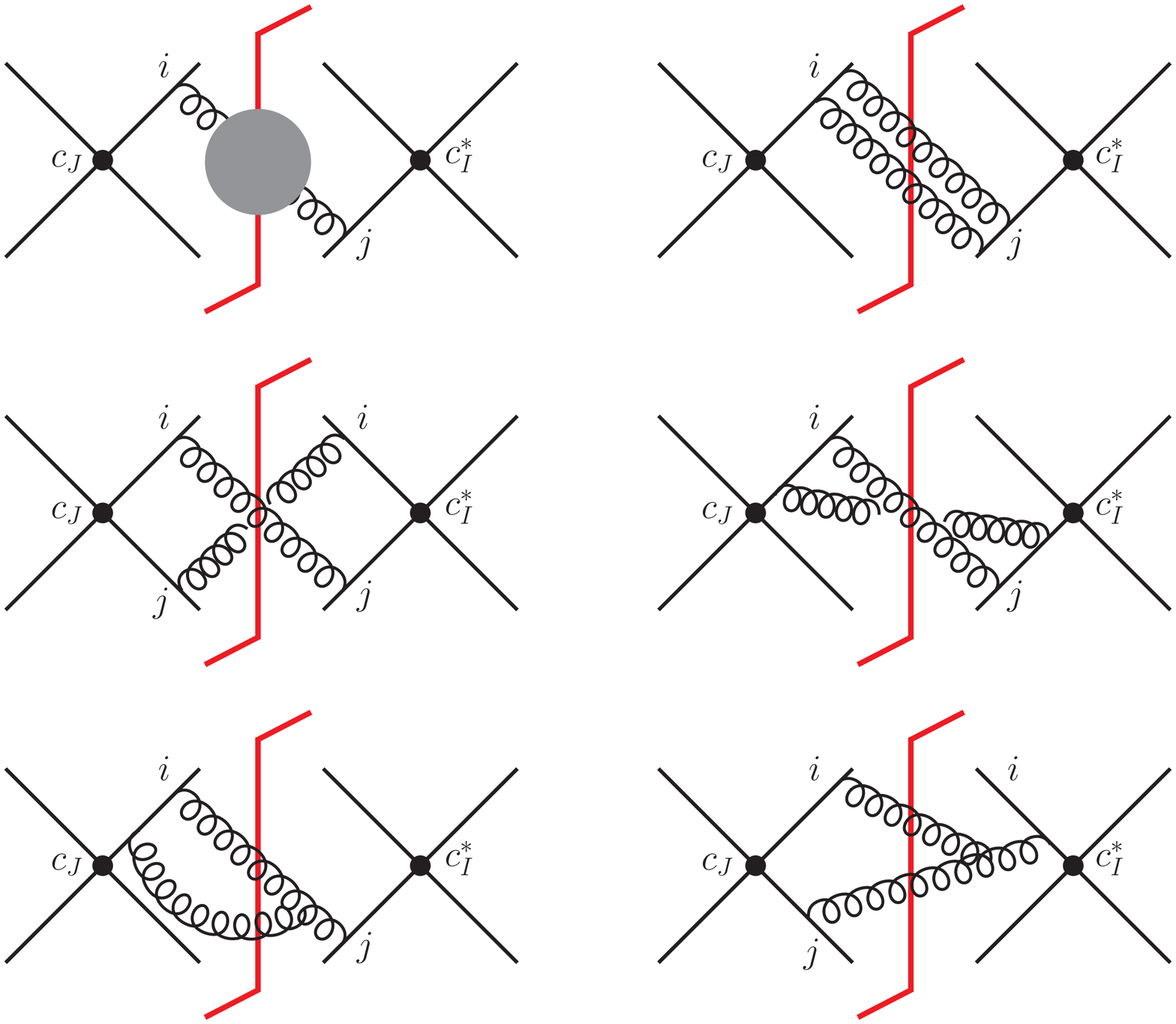}
\includegraphics[width=0.6\textwidth]{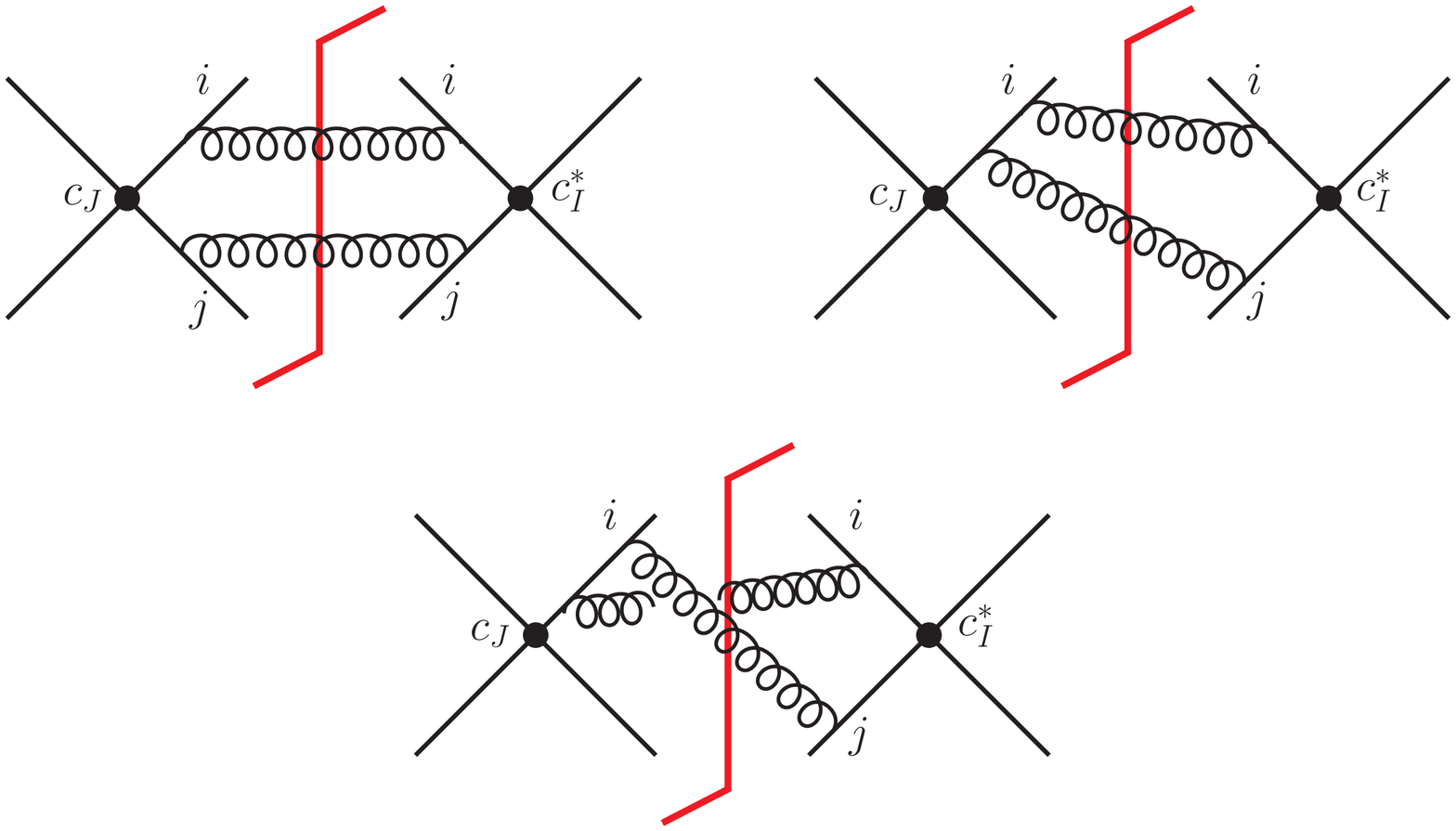}
\end{center}
\vspace{-2ex}
\caption{Two-Wilson-line double-real diagrams contributing to the NNLO soft function.}
\label{fig:NNLORRtwoWilsonLine}
\end{figure}

The Feynman diagrams involving only one Wilson line along the vector $v_i$ are depicted in Figure~\ref{fig:NNLORRoneWilsonLine}. It is clear that such diagrams must be proportional to $v_i^2$, and therefore vanish if $v_i$ is light-like. The shaded-blob in the first diagram denotes loops of quarks, gluons and ghosts (we work in the Feynman gauge). Similarly, Figure~\ref{fig:NNLORRtwoWilsonLine} shows the diagrams involving two Wilson lines in the directions $v_i$ and $v_j$. The integrals coming from these diagrams can be classified into two integral families. The first family is defined by the following set of 6 propagators:
\begin{align}
\label{eq:family1}
\{ (k_1+k_2)^2, \, v_1 \cdot k_2, \, v_1 \cdot (k_1+k_2), \, v_2 \cdot k_1, \, v_3 \cdot k_1, \, v_3 \cdot (k_1+k_2) \} \, .
\end{align}
The corresponding integrals have the form
\begin{align}
\label{eq:F1}
F^{(1)}_{a_1,a_2,a_3,a_4,a_5,a_6} \equiv \int [dk_1] \, [dk_2] \, \delta \big( \omega - v_0 \cdot (k_1+k_2) \big) \, \prod_{i=1}^6 (D_i)^{-a_i} \, ,
\end{align}
where $D_i$ refers to the propagators in Eq.~(\ref{eq:family1}). We again feed all the integrals into \texttt{Reduze2} and \texttt{FIRE5}, which reduce them to 14 master integrals. We collect these master integrals into a vector
\begin{align}
\vec{f}^{(1)}(\epsilon,\beta,y,\omega) \equiv \big( &F^{(1)}_{0,0,0,0,0,0}, F^{(1)}_{1,0,0,0,-1,2}, F^{(1)}_{0,0,0,0,1,0}, F^{(1)}_{0,0,0,0,1,1}, F^{(1)}_{0,1,0,0,0,2}, \nonumber
\\
&F^{(1)}_{0,0,1,0,0,2}, F^{(1)}_{0,1,0,0,1,1}, F^{(1)}_{1,1,0,0,1,-1}, F^{(1)}_{1,1,-1,0,1,0}, F^{(1)}_{1,1,0,0,1,0}, \nonumber
\\
&F^{(1)}_{1,0,1,0,1,0}, F^{(1)}_{0,0,1,0,1,0}, F^{(1)}_{0,1,1,0,1,1}, F^{(1)}_{1,1,0,1,0,0} \big)^{\mathsf{T}} \, .
\end{align}
We transform the above master integrals to a ``canonical basis'' via a linear transformation $\vec{g}^{(1)}(\epsilon,\beta,y) = \hat{T}^{(1)}(\epsilon,\beta,y,\omega) \, \vec{f}^{(1)}(\epsilon,\beta,y,\omega)$, where the transformation matrix $\hat{T}^{(1)}$ is diagonal with its diagonal entries given by
\begin{align}
\hat{T}^{(1)}(\epsilon,\beta,y,\omega) = &\frac{8 \, \Gamma(1-4\epsilon)}{\pi^{2-2\epsilon} \, \omega^{3-4\epsilon} \, \Gamma^2(1-\epsilon)} \nonumber
\\
\times \diag \Big\{ &(1-2\epsilon)(1-4\epsilon)(3-4\epsilon ), \, \epsilon^2(1-2\epsilon )\omega^3\beta, \, \epsilon(1-2\epsilon)(1-4\epsilon)\omega\beta, \nonumber
\\
&\epsilon^2(1-4\epsilon)\omega^2\beta^2, \, \epsilon^2\omega^3\beta^2, \, \epsilon(1-2\epsilon)\omega^3\beta^2, \, \epsilon^3\omega^3\beta(1-\beta y), \, \epsilon^3\omega^3, \nonumber
\\
&\epsilon^3\omega^3\beta, \, \epsilon^3\omega^4(1-\beta y), \, \epsilon^3\omega^4(1-\beta y), \, \epsilon^2(1-4\epsilon)\omega^2(1+\beta y), \nonumber
\\
&\epsilon^3\omega^4(1-\beta y)^2, \, \epsilon^3\omega^4 \Big\} \, .
\end{align}
The transformed vector of master integrals $\vec{g}^{(1)}$ satisfies the differential equation
\begin{align}
\label{eq:diffg1}
\partial_{\beta}\vec{g}^{(1)}(\epsilon,\beta,y) = \epsilon \left( -\frac{\hat{a}^{(1)}}{\beta-1} + \frac{\hat{b}^{(1)}}{\beta} + \frac{\hat{c}^{(1)}}{\beta+1} + \frac{\hat{d}^{(1)}}{\beta+1/y} - \frac{\hat{e}^{(1)}}{\beta-1/y} \right) \vec{g}^{(1)}(\epsilon,\beta,y) \, ,
\end{align}
where $\hat{a}^{(1)}$, $\hat{b}^{(1)}$, $\hat{c}^{(1)}$, $\hat{d}^{(1)}$ and $\hat{e}^{(1)}$ are $14 \times 14$ matrices with matrix elements independent of $\epsilon$ and $\beta$. 

We now turn to the second integral family in the one- or two-Wilson-line diagrams. It is defined by the set of propagators
\begin{align}
\label{eq:family2}
\{ (k_1+k_2)^2, \, v_1\cdot k_1, \, v_1 \cdot (k_1+k_2), \, v_4 \cdot k_1, \, v_3 \cdot k_2, \, v_3 \cdot (k_1+k_2) \} \, .
\end{align}
We denote the corresponding integrals as $F^{(2)}_{a_1,a_2,a_3,a_4,a_5,a_6}$, defined similar to Eq.~(\ref{eq:F1}), but with the propagators $D_i$ chosen from the above set (\ref{eq:family2}). The master integrals in this family can be chosen as
\begin{align}
\vec{f}^{(2)}(\epsilon,\beta,y,\omega) \equiv \big( &F^{(2)}_{0,0,0,0,0,0}, F^{(2)}_{1,0,0,0,-1,2}, F^{(2)}_{0,0,0,0,1,0}, F^{(2)}_{0,0,0,0,1,1}, F^{(2)}_{0,0,0,1,0,0}, \nonumber
\\
&F^{(2)}_{0,0,0,1,1,0}, F^{(2)}_{1,0,0,1,1,0}, F^{(2)}_{0,0,0,1,1,1}, F^{(2)}_{1,0,0,1,0,1} \big)^{\mathsf{T}} \, ,
\end{align}
which are transformed into a canonical basis $\vec{g}^{(2)}(\epsilon,\beta,y)$ by the following transformation matrix
\begin{align}
\hat{T}^{(2)}(\epsilon,\beta,y,\omega) = &\frac{8 \, \Gamma(1-4\epsilon)}{\pi^{2-2\epsilon} \, \omega^{3-4\epsilon} \, \Gamma^2(1-\epsilon)} \nonumber
  \\
\times \diag \Big\{ &(1-2\epsilon)(1-4\epsilon)(3-4\epsilon ), \, \epsilon^2(1-2\epsilon)\omega^3\beta, \, \epsilon(1-2\epsilon)(1-4\epsilon)\omega\beta, \nonumber
\\
&\epsilon^2(1-4\epsilon)\omega^2\beta^2, \, \epsilon(1-2\epsilon)(1-4\epsilon)\omega\beta, \, \epsilon^2(1-4\epsilon)\omega^2\beta^2, \nonumber
\\
&\epsilon^3\omega^4\beta, \, \epsilon^3\omega^3\beta^2, \epsilon^3\omega^4\beta \Big\} \, .
\end{align}
The differential equation satisfied by $\vec{g}^{(2)}$ is given by
\begin{align}
\label{eq:diffg2}
\partial_{\beta}\vec{g}^{(2)}(\epsilon,\beta,y) = \epsilon \left( -\frac{\hat{a}^{(2)}}{\beta-1} + \frac{\hat{b}^{(2)}}{\beta} + \frac{\hat{c}^{(2)}}{\beta+1} \right) \vec{g}^{(2)}(\epsilon,\beta,y) \, ,
\end{align}
with $\hat{a}^{(2)}$, $\hat{b}^{(2)}$ and $\hat{c}^{(2)}$ being $9 \times 9$ constant matrices.

In order to solve the differential equations (\ref{eq:diffg1}) and (\ref{eq:diffg2}), we also need the boundary conditions at some value of $\beta$. Similar to the NLO case, we again choose the point $\beta = 0$ as the boundary, where only 7 of the integrals in $\vec{g}^{(1)}$ and $\vec{g}^{(2)}$ are non-vanishing. Some of the boundary conditions are related to each other, similar to (\ref{eq:boundaryrelation}). The independent ones are given by
\begin{align}
\label{eq:NNLORRboundary}
g_1^{(1)}(\epsilon, 0, y) = g_1^{(2)}(\epsilon, 0, y) &= \frac{4 \, \Gamma(4-4\epsilon)}{\pi^{2-2\epsilon} \, \omega^{3-4\epsilon} \, \Gamma^2(1-\epsilon)} \, \int [dk_1] \, [dk_2] \, \delta \big( \omega - v_0 \cdot (k_1+k_2) \big) = 1 \, , \nonumber
\\
g_{12}^{(1)}(\epsilon, 0, y) &= \frac{8 \, \Gamma(2-4\epsilon)}{\pi^{2-2\epsilon} \, \omega^{1-4\epsilon} \, \Gamma^2(-\epsilon)} \, \int [dk_1] \, [dk_2] \, \frac{\delta \big( \omega - v_0 \cdot (k_1+k_2) \big)}{v_1 \cdot (k_1+k_2) \; v_3 \cdot k_1} \, , \nonumber
\\
&= \frac{2\pi^2}{3} \, \epsilon^2 + \frac{84\zeta_3}{3} \, \epsilon^3 + \frac{4\pi^4}{3} \, \epsilon^4 + \mathcal{O}(\epsilon^5) \, , \nonumber
\\
g_{14}^{(1)}(\epsilon, 0, y) &= \frac{8 \, \epsilon \, \Gamma(1-4\epsilon) \, \omega^{1+4\epsilon}}{\pi^{2-2\epsilon} \, \Gamma^2(-\epsilon)} \, \int [dk_1] \, [dk_2] \, \frac{\delta \big( \omega - v_0 \cdot (k_1+k_2) \big)}{(k_1+k_2)^2 \; v_1 \cdot k_2 \; v_2 \cdot k_1} \, , \nonumber
\\
&= -\frac{4 \, \Gamma(-2 \epsilon) \, \Gamma(1-2\epsilon)}{\Gamma(1-\epsilon) \, \Gamma (-3\epsilon)} \, _3F_2(-\epsilon, -\epsilon, -\epsilon; 1-\epsilon, -3 \epsilon; 1) \, .
\end{align}
The hypergeometric function appearing in the above formula can be expanded in $\epsilon$ with the help of the program package \texttt{HypExp} \cite{Huber:2007dx}. The differential equations of $\vec{g}^{(1)}$ and $\vec{g}^{(2)}$ can then be solved order-by-order in $\epsilon$ in terms of the GPLs, similar to the NLO case (\ref{eq:nloDEQ2}).

\subsubsection{Three-Wilson-line diagrams}
\label{sec:rr3}

We now turn to the three-Wilson-line diagrams. In the calculation of the two-loop anomalous dimensions and infrared singularities \cite{Ferroglia:2009ep, Ferroglia:2009ii}, it was found that the three-Wilson-line diagrams are the most complicated ones. They give rise to the three-parton functions $F_1$ and $f_2$ in Eq.~(\ref{eq:gammaS}). It is therefore highly interesting to see how these complications appear in the calculation of the NNLO soft function. As will be clear below, the genuine three-parton correlations only arise from the virtual-real contributions, and the double-real three-parton contributions can always be expressed as convolutions of two NLO integrals. This can be understood since the soft function is a Hermitian matrix from its definition (\ref{eq:Smom}). On the other hand, the genuine three-parton contributions (such as the functions $F_1$ and $f_2$) multiply the anti-Hermitian color factor $if^{abc}\bm{T}^a_i\bm{T}^b_j\bm{T}^c_k$. Therefore the soft function can only receive contributions from the imaginary parts of the three-parton integrals, which are only present in the virtual-real diagrams but not in the double-real diagrams.

\begin{figure}[t!]
\centering
\includegraphics[width=0.6\textwidth]{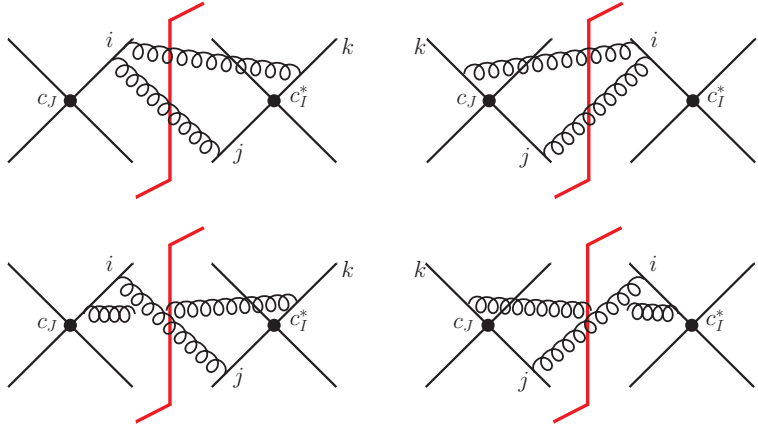}
\vspace{-2ex}
\caption{A set of three-Wilson-line double-real diagrams adding up to a convolution of two NLO integrals.}
\label{fig:NNLORRthreeWilsonLine1}
\end{figure}

\begin{figure}[t!]
\centering
\includegraphics[width=0.6\textwidth]{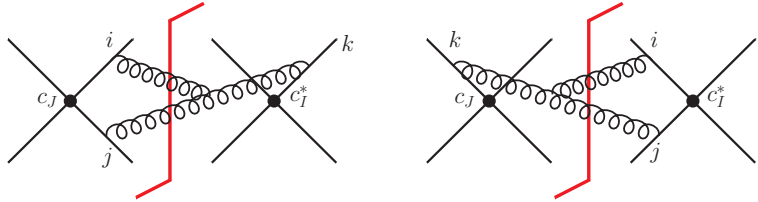}
\vspace{-2ex}
\caption{A pair of three-Wilson-line double-real diagrams adding up to zero.}
\label{fig:NNLORRthreeWilsonLine2}
\end{figure}

More practically, one can carry out an analysis similar to that in \cite{Ferroglia:2012uy} (done for the massless soft function). In that paper, it was demonstrated that the three-Wilson-line integrals in the double-real contributions can be combined into convolutions of NLO integrals. The same applies to the massive soft function. To see this, consider for example the set of diagrams shown in Figure~\ref{fig:NNLORRthreeWilsonLine1}. Each of these diagrams gives rise to rather complicated integrals. However, the sum of the four diagrams leads to the simple integral
\begin{align}
\int [dk_1] \, [dk_2] \, \frac{ v_i \cdot v_j \; v_i \cdot v_k \; \delta \big( \omega - v_0 \cdot (k_1+k_2) \big) }{ v_i \cdot k_1 \; v_j \cdot k_1 \; v_i \cdot k_2 \; v_k \cdot k_2 } \, ,
\end{align}
which is obviously a convolution of two NLO integrals. This fact does not depend on whether $v_i$, $v_j$ and $v_k$ are light-like or time-like, and therefore applies equally well to the massless and the massive soft functions. Another example is the two diagrams involving the three-gluon vertex, shown in Figure~\ref{fig:NNLORRthreeWilsonLine2}. As demonstrated in \cite{Ferroglia:2012uy}, they add up to zero due to the color structure, irrespective of the nature of the Wilson lines involved.

\subsection{Virtual-real contributions}

In this subsection, we present the calculation of the virtual-real contributions. As discussed in Section~\ref{sec:rr3}, the virtual-real diagrams contain genuine three parton correlations. In particular, the scale-dependent part of their contributions involves the complicated function $f_2$ in the anomalous dimension matrix (\ref{eq:gammaS}) calculated in \cite{Ferroglia:2009ep, Ferroglia:2009ii}. It can be expected that the calculation of the scale-independent pieces will be more involved. It was not known at all whether or not they can be written in terms of GPLs. In our explicit calculations, we find that a canonical basis can be constructed and therefore all the master integrals for the virtual-real contributions can be solved iteratively as GPLs to all orders in the dimensional regulator $\epsilon$.

\begin{figure}[t!]
\begin{center}
\includegraphics[width=0.6\textwidth]{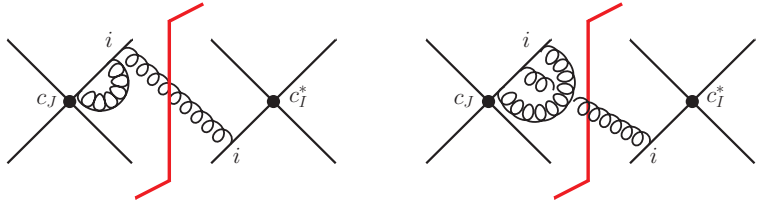}
\end{center}
\vspace{-2ex}
\caption{One-Wilson-line virtual-real diagrams contributing to the NNLO soft function.}
\label{fig:NNLOVRoneWilsonLine}
\end{figure}

\begin{figure}[t!]
\begin{center}
\includegraphics[width=0.6\textwidth]{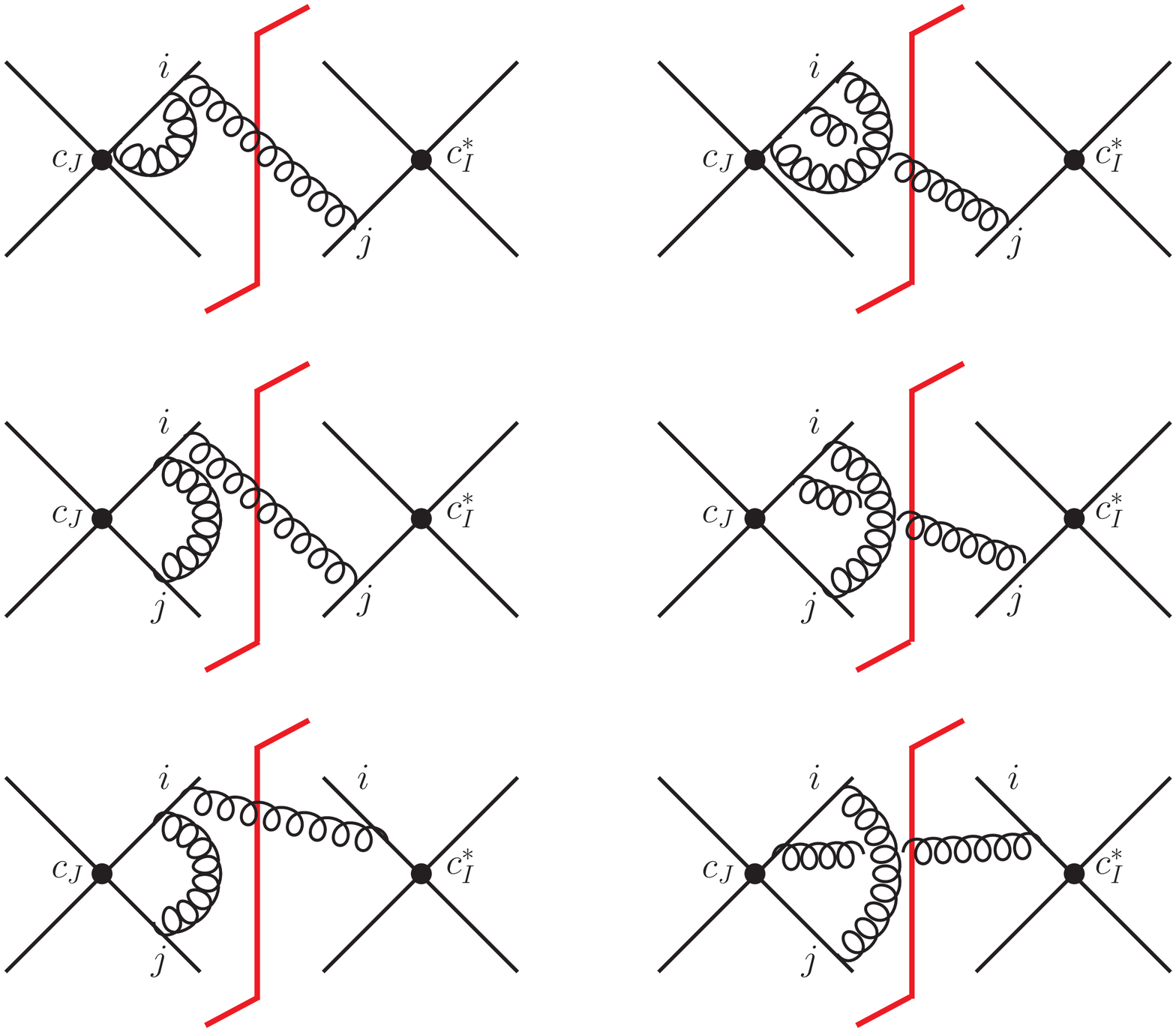}
\includegraphics[width=0.6\textwidth]{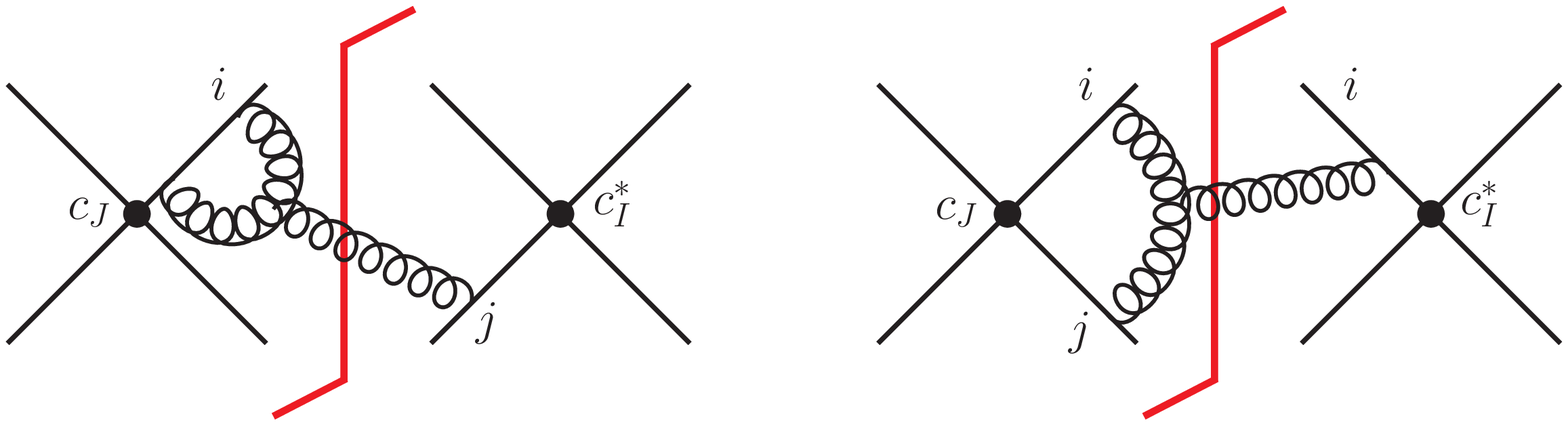}
\end{center}
\vspace{-2ex}
\caption{Two-Wilson-line virtual-real diagrams contributing to the NNLO soft function.}
\label{fig:NNLOVRtwoWilsonLine}
\end{figure}

\begin{figure}[t!]
\centering
\includegraphics[width=0.6\textwidth]{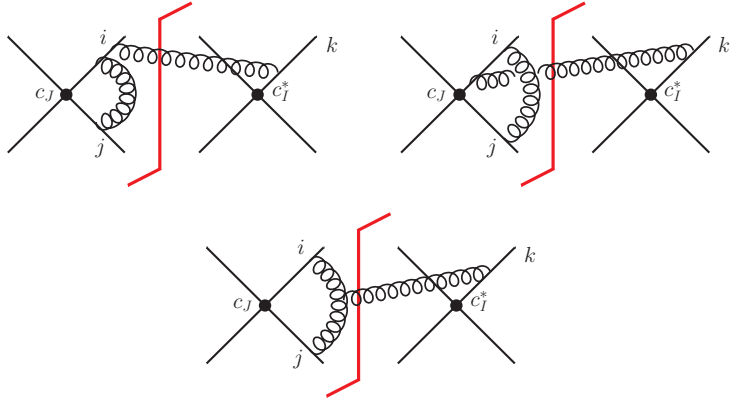}
\caption{Three-Wilson-line virtual-real diagrams.}
\label{fig:NNLOVRthreeWilsonLine}
\end{figure}

The one-, two- and three-Wilson-line virtual-real diagrams are shown in Figure~\ref{fig:NNLOVRoneWilsonLine}, \ref{fig:NNLOVRtwoWilsonLine} and \ref{fig:NNLOVRthreeWilsonLine}, respectively. It is easy to understand that the four Wilson line diagrams always lead to scaleless integrals in dimensional regularization and we don't need to consider them. We write a generic virtual-real integral in the form
\begin{align}
\int d^dk \, d^dl \, \delta \big( \omega - v_0 \cdot k \big) \, \mathrm{Disc} \big[ (k^2)^{-a_1} \big] \, \mathcal{F}(\{v\},k,l) \, ,
\end{align}
where $l$ is the loop momentum, and we have extracted the delta-function from the phase-space measure $[dk]$ into
\begin{align}
\mathrm{Disc} \big[ (k^2)^{-a_1} \big] \equiv \frac{1}{2\pi i} \left[ (k^2+i0)^{-a_1} - (k^2-i0)^{-a_1} \right] .
\end{align}
We find this convenient since we will use the method of differential equation to raise the power $a_1$. The virtual-real integrals can again be classified into two integral families. The first family is defined by the following set of 7 propagators $D_i$ ($i=2,\ldots,8$):
\begin{align}
\label{eq:family3}
\{ l^2, \, (k+l)^2, \, v_1 \cdot k, \, v_1 \cdot l, \, v_2 \cdot (k+l), \, v_3 \cdot k, \, v_3 \cdot l \} \, .
\end{align}
The corresponding integrals can be expressed as
\begin{align}
F^{(3)}_{a_1,a_2,a_3,a_4,a_5,a_6,a_7,a_8} = \int d^dk \, d^dl \, \delta \big( \omega - v_0 \cdot k \big) \, \mathrm{Disc} \big[ (k^2)^{-a_1} \big] \prod_{i=2}^8 (D_i)^{-a_i} \, .
\end{align}
We choose the master integrals in this family to be
\begin{align}
\vec{f}^{(3)}(\epsilon,\beta,y,\omega) \equiv \big( &F_{1,1,1,0,0,0,-1,2}^{(3)}, F_{2,0,1,0,0,0,0,2}^{(3)}, F_{1,0,1,1,0,0,0,2}^{(3)}, F_{1,1,0,0,0,1,0,1}^{(3)}, F_{1,1,1,0,0,1,-1,1}^{(3)}, \nonumber
\\
&F_{1,1,1,-1,0,1,0,1}^{(3)}, F_{1,1,1,0,0,1,0,1}^{(3)}, F_{1,1,1,1,-1,1,0,1}^{(3)}, F_{1,1,1,1,0,1,0,1}^{(3)}, F_{1,1,1,0,1,1,0,0}^{(3)}, \nonumber
\\
&F_{1,1,1,-1,1,1,1,0}^{(3)}, F_{1,1,1,0,1,1,1,0}^{(3)}, F_{1,1,1,-1,1,1,1,-1}^{(3)}, F_{1,1,0,0,0,1,1,2}^{(3)}, F_{1,1,0,0,0,1,1,1}^{(3)} \big)^{\mathsf{T}} \, .
\end{align}
The transformation matrix
\begin{align}
\hat{T}^{(3)}(\epsilon, \beta, y, \omega) &= \frac{ i e^{-2i\pi\epsilon} \, \omega^{4\epsilon} \, \Gamma(-2\epsilon) }{ \pi^{3-2\epsilon} \, \Gamma^2(-\epsilon) \, \Gamma(1+2\epsilon) } \, \hat{T}'^{(3)}(\epsilon, \beta, y, \omega)
\end{align}
takes the master integrals into a canonical basis $\vec{g}^{(3)}(\epsilon,\beta,y)$, which satisfies the differential equation
\begin{align}
\label{eq:diffg3}
\partial_{\beta}\vec{g}^{(3)}(\epsilon,\beta,y) = \epsilon \left( -\frac{\hat{a}^{(3)}}{\beta-1} + \frac{\hat{b}^{(3)}}{\beta} + \frac{\hat{c}^{(3)}}{\beta+1} - \frac{\hat{d}^{(3)}}{\beta-1/y} + \frac{\hat{e}^{(3)}}{\beta+1/y} \right) \vec{g}^{(3)}(\epsilon,\beta,y) \, .
\end{align}
The second integral family in the virtual-real contributions is defined by the propagators
\begin{align}
\label{eq:family4}
\{ l^2, \, (k+l)^2, \, v_1 \cdot k, \, v_1 \cdot l, \, v_4 \cdot (k+l), \, v_3 \cdot k, \, v_3 \cdot l \} \, .
\end{align}
We choose the master integrals to be
\begin{align}
\vec{f}^{(4)}(\epsilon,\beta,y,\omega) \equiv \big( &F_{1,1,1,0,0,0,0,1}^{(4)}, F_{2,0,1,0,0,0,0,2}^{(4)}, F_{1,2,0,0,0,1,1,0}^{(4)}, F_{1,1,0,1,0,2,0,0}^{(4)}, F_{1,0,1,1,0,0,0,2}^{(4)}, \nonumber
\\
&F_{2,1,0,0,0,1,0,1}^{(4)}, F_{1,1,0,0,0,2,0,1}^{(4)}, F_{1,1,0,0,0,1,1,1}^{(4)}, F_{1,0,1,1,0,1,0,1}^{(4)}, F_{1,1,1,0,0,1,0,1}^{(4)}, \nonumber
\\
&F_{1,1,0,1,0,1,0,1}^{(4)}, F_{1,1,1,0,1,1,0,0}^{(4)}, F_{1,1,1,-1,1,1,0,0}^{(4)}, F_{1,1,1,0,1,1,-1,0}^{(4)}, F_{1,1,1,-1,1,1,1,0}^{(4)}, \nonumber
\\
&F_{1,1,1,0,1,1,1,0}^{(4)}, F_{1,1,1,0,1,1,1,-1}^{(4)}, F_{1,1,1,1,-1,1,0,1}^{(4)}, F_{1,1,1,1,0,1,-1,1}^{(4)}, F_{1,1,1,1,0,1,0,1}^{(4)}, \nonumber
\\
&F_{1,1,1,1,-1,1,-1,1}^{(4)} \big)^{\mathsf{T}} \, ,
\end{align}
with the transformation matrix
\begin{align}
\hat{T}^{(4)}(\epsilon, \beta, y, \omega) &= \frac{ i e^{-2 i\pi\epsilon} \,  \omega^{4\epsilon} \, \Gamma(-2\epsilon) }{ \pi^{3-2\epsilon} \, \Gamma^2(-\epsilon) \, \Gamma(1+2\epsilon) } \, \hat{T}'^{(4)}(\epsilon, \beta, y, \omega) \, .
\end{align}
The resulting canonical basis $\vec{g}^{(4)}(\epsilon,\beta,y)$ satisfies the differential equation
\begin{align}
\label{eq:diffg4}
\partial_{\beta}\vec{g}^{(4)}(\epsilon,\beta,y) = \epsilon \left( -\frac{\hat{a}^{(4)}}{\beta-1} + \frac{\hat{b}^{(4)}}{\beta} + \frac{\hat{c}^{(4)}}{\beta+1} - \frac{\hat{d}^{(4)}}{\beta-1/y} + \frac{\hat{e}^{(4)}}{\beta+1/y} \right) \vec{g}^{(4)}(\epsilon,\beta,y) \, .
\end{align}
The $15 \times 15$ matrices $\hat{T}'^{(3)}$, $\hat{a}^{(3)}$, $\hat{b}^{(3)}$, $\hat{c}^{(3)}$, $\hat{d}^{(3)}$, $\hat{e}^{(3)}$ as well as the $21 \times 21$ matrices $\hat{T}'^{(4)}$, $\hat{a}^{(4)}$, $\hat{b}^{(4)}$, $\hat{c}^{(4)}$, $\hat{d}^{(4)}$, $\hat{e}^{(4)}$ are non-diagonal with lengthy expressions. We choose to give them in an electronic file attached to the arXiv submission, together with the matrices $\hat{a}^{(1)}$, $\hat{b}^{(1)}$, $\hat{c}^{(1)}$, $\hat{d}^{(1)}$, $\hat{e}^{(1)}$, $\hat{a}^{(2)}$, $\hat{b}^{(2)}$ and $\hat{c}^{(2)}$ appearing in the calculation of the double-real contributions.

In order to solve the differential equations (\ref{eq:diffg3}) and (\ref{eq:diffg4}), we again need to calculate the boundary conditions at $\beta = 0$. In the virtual-real case, an additional complication arises due to the fact that some of the master integrals are logarithmic divergent in the limit $\beta \to 0$. This happens when the loop integral (involving both $v_3$ and $v_4$) give rise to Coulomb-like singularities of $1/\beta$ which are multiplied by some $\epsilon$-dependent powers of $\beta$. When this is the case, we cannot set $\beta = 0$ before performing the integration, like what we did in the double-real case. Instead, we have to explicitly calculate the $\beta$-dependence of such master integrals (at least their asymptotic form as $\beta \to 0$). Fortunately, this only needs to be done for two of the master integrals, which are
\begin{align}
g^{(4)}_6(\epsilon,\beta,y) &= \frac{ i e^{-2 i \pi\epsilon} \, \Gamma(-2\epsilon) \, \omega^{-1+4\epsilon} \, \beta }{ 2 \pi^{3-2\epsilon} \, \Gamma^2(-\epsilon) \, \Gamma(2\epsilon) } \int [dk] \, d^dl \, \frac{\delta(\omega - v_0 \cdot k)}{(l^2+i0) \; (v_3 \cdot l + i0) \; [-v_4 \cdot (k+l) + i0] } \nonumber
\\
&\hspace{15em} \times \left[ \frac{\omega}{-v_4 \cdot (k+l) + i0} + 2(1-4\epsilon) \right] ,
\\
g^{(4)}_9(\epsilon,\beta,y) &= -\frac{ i e^{-2 i \pi\epsilon} \, \Gamma(1-2\epsilon) \, \omega^{4\epsilon} \, \beta }{ 4 \pi^{3-2\epsilon} \, \Gamma^2(-\epsilon) \, \Gamma(2\epsilon) } \int [dk] \, d^dl \, \frac{\delta(\omega - v_0 \cdot k)}{(l^2+i0) \; (-v_1 \cdot k) \; (v_3 \cdot l + i0) }  \nonumber
\\
&\hspace{15em} \times \frac{1}{-v_4 \cdot (k+l) + i0} \, .
\end{align}
Here, we explicitly write the imaginary parts of the propagators involving the loop momentum $l$, since the loop integrals over $l$ depend on them. In the following, we will suppress the imaginary parts and the $+i0$ prescription is always understood. The results of the integrals over the loop momentum $l$ can be found in \cite{Bierenbaum:2011gg, Czakon:2018iev}.
The remaining integrations over the momentum of the real gluon $k$ can be carried out in the limit $\beta \to 0$, since no new divergence arises in this limit. The asymptotic behaviors of $g^{(4)}_6$ and $g^{(4)}_9$ near the boundary can then be obtained and are given by
\begin{align}
g^{(4)}_6(\epsilon,\beta \to 0,y) &\approx \frac{ (e^{-2 i\pi\epsilon} - 1) \, \beta^{2\epsilon} \, \Gamma(1-2\epsilon) \, \Gamma(1+\epsilon) }{ 4^{1-2\epsilon} \, \Gamma(1-\epsilon) } \, ,
\\
g^{(4)}_9(\epsilon,\beta \to 0,y) &\approx \frac{ (e^{-2 i\pi\epsilon} - 1) \, \beta^{2\epsilon} \, \Gamma(1-2\epsilon) \, \Gamma(1+\epsilon) }{ 2^{3-4\epsilon} \, \Gamma(1-\epsilon) } \nonumber
\\
&= \frac{1}{2} \, g^{(4)}_6(\epsilon,\beta \to 0,y) \, .
\end{align}
This can be readily expanded in $\epsilon$ to arrive at the boundary conditions. The other independent boundary conditions are easier to obtain and are given by
\begin{align}
\label{eq:NNLOVRboundary1}
g^{(3)}_2(\epsilon, 0, y) &= g^{(4)}_2(\epsilon, 0, y) = -\frac{ i e^{-2i\pi\epsilon } \, (1-2 \epsilon) \, \Gamma(2-2\epsilon) }{ \pi^{3-2\epsilon} \, \omega^{2-4\epsilon} \, \Gamma^2(1-\epsilon) \Gamma(2\epsilon) } \int [dk] \, d^dl \, \frac{\delta(\omega - v_0 \cdot k)}{(k+l)^2 \; v_3 \cdot l} \, \nonumber
\\
&=1 \, , \nonumber
\\
g^{(3)}_4(\epsilon, 0, y) &= \frac{ i e^{-2 i \pi\epsilon } \, (1-4\epsilon) \, \Gamma (-2\epsilon ) }{ 2 \pi^{3-2\epsilon} \, \omega^{1-4\epsilon} \, \Gamma^2(-\epsilon) \, \Gamma (2\epsilon)} \int [dk] \, d^dl \, \frac{\delta(\omega - v_0 \cdot k)}{l^2 \; [ - v_2 \cdot (k+l) ] \; v_3 \cdot l} \, \nonumber
\\
&= - \frac{ e^{-2 i \pi\epsilon } \, \Gamma(1-3\epsilon) \, \Gamma^2(-2\epsilon) \, \Gamma(1+\epsilon) }{ \Gamma(1-4\epsilon) \, \Gamma^2(-\epsilon) } \, , \nonumber
\\
g^{(3)}_9(\epsilon, 0, y) &= \frac{ i e^{-2 i \pi\epsilon } \, \omega^{1+4\epsilon} \, \Gamma (1-2\epsilon ) }{ 4 \pi^{3-2\epsilon}  \, \Gamma^2(-\epsilon) \, \Gamma (2\epsilon)} \int [dk] \, d^dl \, \frac{\delta(\omega - v_0 \cdot k)}{l^2 \; (k+l)^2 \; [ - v_2 \cdot (k+l) ] \; v_3 \cdot l}  \nonumber
\\
&\hspace{15em} \times \left[ \frac{\omega}{-v_1 \cdot k} + 2 \right] \nonumber
\\
&= \frac{1}{3} - \frac{i\pi}{6} \epsilon - \frac{\pi^2}{6} \epsilon^2 + \left( \frac{38\zeta_3}{9} + \frac{4i\pi^3}{3} \right) \epsilon^3 - \left( \frac{209\pi^4}{240} + \frac{79i\pi\zeta_3}{9} \right) \epsilon^4 + \mathcal{O}(\epsilon^5)  \, , \nonumber
\\
g^{(3)}_{10}(\epsilon, 0, y) &= \frac{ i e^{-2 i\pi\epsilon } \, \omega^{1+4\epsilon} \, \Gamma(1-2\epsilon) }{ 4\pi^{3-2\epsilon} \, \Gamma^2(-\epsilon) \, \Gamma(2\epsilon)}  \int [dk] \, d^dl \, \frac{ \delta(\omega - v_0 \cdot k) }{ l^2 \; (k+l)^2 \; [-v_2 \cdot (k+l)] \; v_2 \cdot l} \, \nonumber
 \\
 &= \frac{ e^{-3 i \pi \epsilon } \, \Gamma^2(1-2\epsilon) \, \Gamma^2(1+\epsilon) }{ 2 \, \Gamma(1-4\epsilon) \, \Gamma(1+2\epsilon) } \, .
\end{align}

With the above results, it is straightforward to derive the virtual-real contributions to the NNLO soft function. It should be emphasized again that the three-Wilson-line diagrams give non-vanishing contributions, which are necessary to give the correct pole structure in accordance with the anomalous dimension (\ref{eq:gammaS}). On the other hand, these contributions are proportional to the anti-Hermitian color factor $if^{abc}\bm{T}^a_i\bm{T}^b_j\bm{T}^c_k$, and only enter the off-diagonal entries of the soft function in the singlet-octet basis (\ref{eq:colorbasis}). They therefore do not appear at the level of the NNLO cross section, but will enter the resummation formula which encodes higher order effects beyond the NNLO.

\subsection{The bare soft function in the moment space}

Assembling all the ingredients in the last two subsections, we obtain the NNLO momentum-space bare soft function $\bm{S}_{\text{bare}}^{(2)}(\omega,\beta,y)$. It is written in terms of star-distributions in $\omega$. For later convenience, it is useful to transform the soft function to the moment space using a Laplace transform (\ref{eq:laplace}). This can be most easily done by observing that the $\omega$-dependence of the NNLO bare soft function comes from an overall factor $\omega^{-1-4\epsilon}$, and
\begin{align}
\int_0^\infty d\omega \, \exp \! \left( -\frac{\omega}{\Lambda e^{\gamma_E}} \right) \mu^{4\epsilon} \, \omega^{-1-4\epsilon} = e^{-4\epsilon\gamma_E} \, \Gamma(-4\epsilon) \left( \frac{\mu^2}{\Lambda^2} \right)^{2\epsilon} \, .
\end{align}
A similar transformation rule can be derived for the NLO bare soft function. The resulting moment-space soft function can then be written as a function of $\Lambda$:
\begin{align}
\tilde{\bm{s}}_{\text{bare}}(\Lambda,\beta,y) = \int_0^\infty d\omega \, \exp \! \left( -\frac{\omega}{\Lambda e^{\gamma_E}} \right) \bm{S}_{\text{bare}}(\omega,\beta,y) \, .
\end{align}
Similar to the momentum-space soft function, we define the matrix elements of the moment-space soft function as
\begin{align}
\tilde{\bm{s}}_{IJ}(\Lambda,\beta,y) \equiv \braket{c_I | \tilde{\bm{s}}(\Lambda,\beta,y) | c_J} \, ,
\end{align}
where the color basis is chosen as in Eq.~(\ref{eq:colorbasis}). This matrix-valued moment-space soft function will be the main object to be studied in the following.

\section{Renormalized soft function}
\label{sec:ren}

\subsection{Anomalous dimensions and renormalization constants}

The bare soft functions $\tilde{\bm{s}}_{\text{bare}}$ we just calculated contain ultraviolet divergences. As discussed in Section~\ref{sec:form}, these can be renormalized in the form
\begin{align}
\label{eq:sIJren}
\tilde{\bm{s}}(L,\beta,\cos\theta,\mu) = \lim_{\epsilon \to 0} \bm{Z}_s^{\dagger}(L,\beta,\cos\theta,\mu) \, \tilde{\bm{s}}_{\text{bare}}(\Lambda,\beta,\cos\theta) \, \bm{Z}_s(L,\beta,\cos\theta,\mu) \, ,
\end{align}
where the renormalization matrix $\bm{Z}_s$ can be constructed from the anomalous dimension matrix $\bm{\Gamma}_s(L,\beta,\cos\theta,\mu)$. Taking the matrix elements of the above formula in the color basis (\ref{eq:colorbasis}), we arrive at
\begin{align}
\label{eq:sIJren1}
\tilde{\bm{s}}_{IJ}(L,\beta,\cos\theta,\mu) = \lim_{\epsilon \to 0} \sum_{M,N} \frac{ \braket{c_I | \bm{Z}_s^{\dagger}(L,\beta,\cos\theta,\mu) | c_M} }{ \braket{c_M | c_M} } \, \tilde{\bm{s}}^{\text{bare}}_{MN}(\Lambda,\beta,\cos\theta) \, \frac{ \braket{c_N | \bm{Z}_s(L,\beta,\cos\theta,\mu) | c_J} }{ \braket{c_N | c_N} } \, .
\end{align}
This motivates us to define the matrix elements of the renormalization factor $\bm{Z}_s$ (similar for the anomalous dimension $\bm{\Gamma}_s$) as
\begin{align}
\bm{Z}_{IJ}(L,\beta,\cos\theta,\mu) \equiv \frac{1}{\braket{c_I|c_I}} \braket{c_I | \bm{Z}_s(L,\beta,\cos\theta,\mu) | c_J} \, .
\end{align}
Eq.~(\ref{eq:sIJren1}) can then be written as
\begin{align}
\label{eq:sIJren2}
\tilde{\bm{s}}_{IJ}(L,\beta,\cos\theta,\mu) = \lim_{\epsilon \to 0} \sum_{M,N} \bm{Z}^\dagger_{IM}(L,\beta,\cos\theta,\mu) \, \tilde{\bm{s}}^{\text{bare}}_{MN}(\Lambda,\beta,\cos\theta) \, \bm{Z}_{NJ}(L,\beta,\cos\theta,\mu) \, .
\end{align}

We now need to construct the renormalization matrix out of the anomalous dimension matrix given in Eq.~(\ref{eq:gammaS}), using the relation (\ref{eq:Zs}). Adopting the singlet-octet basis (\ref{eq:colorbasis}), the explicit matrix forms of $\bm{\Gamma}_s$ in the $q\bar{q}$ channel and the $gg$ channel are given by
\begin{align}
\bm{\Gamma}_s^{q\bar{q}} &= \left[ C_F \, \gamma_{\text{cusp}}(\alpha_s) \left( \ln\frac{\Lambda^2}{\mu^2} - i\pi \right) + C_F \, \gamma_{\text{cusp}}(\beta_{34},\alpha_s) + 2\gamma_s^q(\alpha_s) + 2\gamma^Q(\alpha_s) \right] \bm{1} \nonumber
\\
&\quad + \frac{N}{2} \left[ \gamma_{\text{cusp}}(\alpha_s) \left( \ln\frac{t_1^2}{M^2m_t^2} + i\pi \right) - \gamma_{\text{cusp}}(\beta_{34},\alpha_s) \right]
\begin{pmatrix}
0 & 0
\\
0 & 1
\end{pmatrix}
\nonumber
\\
&\quad + \gamma_{\text{cusp}}(\alpha_s) \, \ln\frac{t_1^2}{u_1^2} \left[
\begin{pmatrix}
0 & \frac{C_F}{2N}
\\
1 & -\frac{1}{N}
\end{pmatrix}
+ \frac{\alpha_s}{4\pi} \, g(\beta_{34})
\begin{pmatrix}
0 & \frac{C_F}{2}
\\
-N & 0
\end{pmatrix}
\right] ,
\end{align}
and
\begin{align}
\bm{\Gamma}_s^{gg} &= \left[ N \, \gamma_{\text{cusp}}(\alpha_s) \left( \ln\frac{\Lambda^2}{\mu^2} - i\pi \right) + C_F \, \gamma_{\text{cusp}}(\beta_{34},\alpha_s) + 2\gamma_s^g(\alpha_s) + 2\gamma^Q(\alpha_s) \right] \bm{1} \nonumber
\\
&\quad + \frac{N}{2} \left[ \gamma_{\text{cusp}}(\alpha_s) \left( \ln\frac{t_1^2}{M^2m_t^2} + i\pi \right) - \gamma_{\text{cusp}}(\beta_{34},\alpha_s) \right]
\begin{pmatrix}
0 & 0 & 0
\\
0 & 1 & 0
\\
0 & 0 & 1
\end{pmatrix}
\nonumber
\\
&\quad + \gamma_{\text{cusp}}(\alpha_s) \, \ln\frac{t_1^2}{u_1^2} \left[
\begin{pmatrix}
0 & \frac{1}{2} & 0
\\
1 & -\frac{N}{4} & \frac{N^2-4}{4N}
\\
0 & \frac{N}{4} & -\frac{N}{4}
\end{pmatrix}
+ \frac{\alpha_s}{4\pi}\,g(\beta_{34})
\begin{pmatrix}
0 & \frac{N}{2} & 0
\\
-N & 0 & 0 \\
0 & 0 & 0
\end{pmatrix}
\right] ,
\end{align}
where the various functions were given in Section~\ref{sec:form}. It should be noted that starting from the NNLO one also needs to renormalize the strong coupling $\alpha_s$.

Inserting the renormalization matrices and the bare soft functions into Eq.~(\ref{eq:sIJren2}), we find that all the poles in $\epsilon$ cancel for both the $q\bar{q}$ channel and the $gg$ channel. This provides a strong check on the correctness of our calculation. We can then safely take the limit $\epsilon \to 0$ and obtain finite soft function matrices $\tilde{\bm{s}}^{q\bar{q}}_{IJ}(L,\beta,\cos\theta,\mu)$ and $\tilde{\bm{s}}^{gg}_{IJ}(L,\beta,\cos\theta,\mu)$. These are the main results of our paper. Since the expressions are rather lengthy, we do not give them here and provide instead an electronic file included in the arXiv submission. For illustration purposes, in the following we list the $\mu$-independent terms proportional to the number of light quarks $N_l$ in the octet-octet entry for the $q\bar{q}$ channel:
\begin{align}
\tilde{\bm{s}}^{q\bar{q},(2)}_{22}(0,\beta,y) \bigg|_{T_FN_l} &= \frac{16 (7\beta^2-126\beta+127)}{243\beta} G_{1} + \frac{8(5\beta^2+90\beta+53)}{81\beta} \big( G_{-1,-1} - G_{-1,1} - 2G_{0,-1} \big) \nonumber
\\
&\hspace{-6em} - \frac{16(7\beta^2+126\beta+127)}{243\beta} G_{-1} + \frac{8(5\beta^2 - 90\beta + 53)}{81\beta} \big( G_{1,-1} - G_{1,1} + 2G_{0,1} \big) \nonumber
\\
&\hspace{-6em} + \frac{8(\beta^2+18\beta+1)}{27\beta} \big( -G_{-1,-1,-1} + G_{-1,-1,1} + 2G_{-1,0,-1} - 2G_{-1,0,1} - G_{-1,1,-1} + G_{-1,1,1} \nonumber
\\
&\hspace{-6em} + 2G_{0,-1,-1} - 2G_{0,-1,1} - 4G_{0,0,-1} \big) + \frac{8(\beta^2-18\beta+1)}{27\beta} \big( 4G_{0,0,1} + 2G_{0,1,-1} - 2G_{0,1,1} \nonumber
\\
&\hspace{-6em} - G_{1,-1,-1} + G_{1,-1,1} + 2G_{1,0,-1} - 2G_{1,0,1} - G_{1,1,-1} + G_{1,1,1} \big) \nonumber
\\
&\hspace{-6em} + \frac{32}{243} \bigg[ 28G_{-1/y} + 98G_{1/y} + 30 \big( 2G_{0,-1/y} + G_{-1/y,-1} + G_{-1/y,1} - 2G_{-1/y,-1/y} \big) \nonumber
\\
&\hspace{-6em} + 105 \big( 2G_{0,1/y} + G_{1/y,-1} + G_{1/y,1} - 2G_{1/y,1/y} \big) + 18 \big( 4G_{0,0,-1/y} + 2G_{0,-1/y,-1} + 2G_{0,-1/y,1} \nonumber
\\
&\hspace{-6em} - 4G_{0,-1/y,-1/y} - G_{-1/y,-1,-1} + G_{-1/y,-1,1} + 2G_{-1/y,0,-1} + 2G_{-1/y,0,1} - 4G_{-1/y,0,-1/y} \nonumber
\\
&\hspace{-6em} + G_{-1/y,1,-1} - G_{-1/y,1,1} - 2G_{-1/y,-1/y,-1} - 2G_{-1/y,-1/y,1} + 4G_{-1/y,-1/y,-1/y} \big) \nonumber
\\
&\hspace{-6em} + 63 \big( 4G_{0,0,1/y} + 2G_{0,1/y,-1} + 2G_{0,1/y,1} - 4G_{0,1/y,1/y} - G_{1/y,-1,-1} + G_{1/y,-1,1} + 2G_{1/y,0,-1} \nonumber
\\
&\hspace{-6em} + 2G_{1/y,0,1} - 4G_{1/y,0,1/y} + G_{1/y,1,-1} - G_{1/y,1,1} - 2G_{1/y,1/y,-1} - 2G_{1/y,1/y,1} + 4G_{1/y,1/y,1/y} \big) \nonumber
\\
&\hspace{-6em} - \frac{332}{3} - \frac{5\pi^2}{2} + 6\zeta_3 \bigg] \, ,
\end{align}
where we have set the number of colors $N_c = 3$ in order to shorten the expression, and defined the abbreviations $G_{a_1,\ldots,a_n} \equiv G(a_1,\ldots,a_n;\beta)$.

\subsection{Validations of the results}

While our results of the NNLO soft functions are novel, it is possible to partially validate them by checking their consistency with some known results in the literature. We have performed 3 checks: 1) that they satisfy RGEs according to the anomalous dimension matrices calculated in \cite{Ferroglia:2009ep, Ferroglia:2009ii}; 2) that they reproduce in the threshold limit the results of \cite{Belitsky:1998tc, Czakon:2013hxa}; 3) that they correctly factorize in the boosted limit according to the factorization formula given in \cite{Ferroglia:2012ku}.

As discussed in Section~\ref{sec:form}, the renormalized soft function satisfies a renormalization group equation
\begin{align}
\label{eq:srge2}
\frac{d}{d\mu} \tilde{\bm{s}}(L,\beta,\cos\theta,\mu) &= - \bm{\Gamma}_s^\dagger(L,\beta,\cos\theta,\mu) \, \tilde{\bm{s}}(L,\beta,\cos\theta,\mu) - \tilde{\bm{s}}(L,\beta,\cos\theta,\mu) \, \bm{\Gamma}_s(L,\beta,\cos\theta,\mu) \, .
\end{align}
This property is closely related to the renormalization in Eq.~(\ref{eq:sIJren}). Given that our result is correctly renormalized, we expect that it should naturally satisfies Eq.~(\ref{eq:srge2}). Nevertheless, we have calculated the left side and the right side of the above equation and indeed find consistency.

In the threshold limit $s \to 4m_t^2$ or $\beta \to 0$, the top quark and the anti-top quark are at rest in the partonic center-of-mass frame. In this case, if the $t\bar{t}$ pair forms a color-singlet state, the soft gluons cannot probe it, and the situation is no different from the Drell-Yan process or the Higgs boson production process. The corresponding element of the soft function matrix then reduce to those calculated in \cite{Belitsky:1998tc}. On the other hand, if the $t\bar{t}$ pair forms a color-octet state, the corresponding matrix element in the threshold limit has been calculated in \cite{Czakon:2013hxa}. We can therefore check the consistency of our result by taking the limit $\beta \to 0$ in our expressions. This is easy to do in our formalism since $\beta = 0$ is essentially the boundary point of the differential equations. One can also directly take the $\beta \to 0$ limit starting from the explicit expressions of $\tilde{\bm{s}}(L,\beta,\cos\theta,\mu)$, which is expressed in terms of GPLs of $\beta$. Using the property that $G(a_1,\ldots,a_n;0) = 0$ unless all the indices are zero, this limit is rather straightforward to obtain. We have done this simple exercise and find that the results are in perfect agreement with those in \cite{Belitsky:1998tc} and \cite{Czakon:2013hxa}. In particular, for the $q\bar{q}$ channel we find the diagonal entries to be
\begin{align}
\frac{\tilde{\bm{s}}^{q\bar{q},(2)}_{11}(L,\beta \to 0, \cos\theta)}{\tilde{\bm{s}}^{q\bar{q},(0)}_{11}} &= \frac{C_F^2}{2} \left( 2L^2 + \frac{\pi^2}{3} \right)^2 + C_F C_A \left[ -\frac{22}{9} L^3 + \left( \frac{134}{9} - \frac{2\pi^2}{3} \right) L^2 \right. \nonumber
\\
&+ \left. \left( -\frac{808}{27} + 28\zeta_3 \right) L + \frac{2428}{81} + \frac{67\pi^2}{54} - \frac{22\zeta_3}{9} - \frac{\pi^4}{3} \right] \nonumber
\\
&+ C_F T_F N_l \left( \frac{8}{9} L^3 - \frac{40}{9} L^2 + \frac{224}{27} L - \frac{656}{81} - \frac{10\pi^2}{27} + \frac{8\zeta_3}{9} \right) ,
\\
\frac{\tilde{\bm{s}}^{q\bar{q},(2)}_{22}(L,\beta \to 0, \cos\theta)}{\tilde{\bm{s}}^{q\bar{q},(0)}_{22}} &= \frac{1}{2} \left[ C_F \left( 2L^2 + \frac{\pi^2}{3} \right) + C_A (-2L + 4) \right]^2 + C_F C_A \left[ -\frac{22}{9} L^3  \right. \nonumber
\\
&+ \left. \left( \frac{134}{9} - \frac{2\pi^2}{3} \right) L^2 + \left( -\frac{808}{27} + 28\zeta_3 \right) L + \frac{2428}{81} + \frac{67\pi^2}{54} - \frac{22\zeta_3}{9} - \frac{\pi^4}{3} \right] \nonumber
\\
&+ C_F T_F N_l \left( \frac{8}{9} L^3 - \frac{40}{9} L^2 + \frac{224}{27} L - \frac{656}{81} - \frac{10\pi^2}{27} + \frac{8\zeta_3}{9} \right) \nonumber
\\
&+ C_A^2 \left[ \frac{11}{3} L^2 + \left( -\frac{230}{9} + \frac{2\pi^2}{3} - 4\zeta_3 \right) L + \frac{1568}{27} + \frac{2\pi^2}{3} - 10\zeta_3 + \frac{13\pi^4}{180} \right] \nonumber
\\
&+ C_A T_F N_l \left( -\frac{4}{3} L^2 + \frac{88}{9} L -\frac{640}{27} \right) .
\end{align}
The results for the $gg$ channel can be simply obtained from the above expressions by the replacement $C_F \to C_A$. 

An interesting subtlety in the above exercise is that the non-diagonal entries of the soft function do not vanish in the limit $\beta \to 0$. Instead, they develop logarithmic divergent behaviors in that limit. These non-vanishing terms arise from the three-Wilson-line virtual-real diagrams and are therefore purely imaginary. For example, we have
\begin{align}
\tilde{\bm{s}}^{q\bar{q},(2)}_{12}(L,\beta \to 0, \cos\theta) &= -2i\pi C_F C_A \left[ L^2 - 4 \left( \ln(4\beta) + 1 \right) L + 2\ln^2(4\beta) + 8\ln(4\beta) + \pi^2 \right] .
\end{align}
This is actually expected since the $f_2$ function in the anomalous dimension matrix has the similar property in the threshold limit (see, e.g., Section 3.4 of \cite{Ferroglia:2009ii}). However, such a behavior cannot be seen from the calculation of \cite{Belitsky:1998tc, Czakon:2013hxa}, and is a novel feature of our results.

Besides the above ``trivial'' checks, a highly non-trivial cross-check of our result is the opposite, boosted limit $\beta \to 1$ or $M^2 \gg 4m_t^2$. In this limit, it was shown in \cite{Ferroglia:2012ku} that the soft function should factorize in the form
\begin{align}
\label{eq:sfac}
\tilde{\bm{s}}_{\text{massive}} \! \left(\ln\frac{\Lambda^2}{\mu^2},\beta,\cos\theta,\mu\right) = \tilde{\bm{s}}_{\text{massless}} \! \left(\ln\frac{\Lambda^2}{\mu^2},\cos\theta,\mu\right) \, \tilde{s}_D^2 \! \left(\ln\frac{m_t^2\Lambda^2}{M^2\mu^2},\mu\right) + \mathcal{O}(m_t^2/M^2) \, ,
\end{align}
where $\tilde{\bm{s}}_{\text{massive}}$ is the massive soft function calculated in this work, $\tilde{\bm{s}}_{\text{massless}}$ is the soft function with top quarks treated as massless (which was calculated at NNLO in \cite{Ferroglia:2012uy}\footnote{It should be noted that \cite{Ferroglia:2012uy} did not calculate the three-Wilson-line virtual-real contributions to the massless soft function $\tilde{\bm{s}}_{\text{massless}}$, which however can be easily extracted from the results in this paper. It should be possible to perform the massless calculation explicitly and compare with the expressions obtained here.}), and $\tilde{s}_D$ is a soft-collinear partonic fragmentation function describing a boosted top quark evolving into a top quark plus soft radiations. The soft-collinear fragmentation function $\tilde{s}_D$ was not directly calculated in the literature. In \cite{Gardi:2005yi, Neubert:2007je}, it was related to the shape function in $B$-meson decays and was extracted from existing result of the latter. In the Appendix~A of \cite{Ferroglia:2012ku}, however, it was found that this result of $\tilde{s}_D$ is inconsistent with other related calculations. To understand this, we note that $\tilde{s}_D$ should satisfy another factorization formula
\begin{align}
\label{eq:Dfac}
\tilde{D}_{t/t}(\Lambda^2/s,m_t,\mu) = C_D(m_t,\mu) \, \tilde{s}_D(L_D,\mu) + \mathcal{O}(\Lambda^2/M^2) \, ,
\end{align}
where $L_D \equiv \ln\big(m_t^2\Lambda^2/M^2\mu^2\big)$ (note that $\tilde{s}_D$ depends on a see-saw scale $m_t\Lambda/M$). In the above formula, $\tilde{D}_{t/t}$ is the full partonic fragmentation function of the top quark in the moment space, and $C_D$ is a hard-collinear matching coefficient. The momentum-space version of $\tilde{D}_{t/t}$ was calculated in \cite{Melnikov:2004bm}. Using this result and the result of $\tilde{s}_D$ from \cite{Gardi:2005yi, Neubert:2007je}, it is possible to extract the form of $C_D$ from Eq.~(\ref{eq:Dfac}). However, the function $C_D$ could also be extracted from the high-energy behavior of the scattering-amplitude involving heavy quarks \cite{Mitov:2006xs}. The main conclusion of the Appendix~A of \cite{Ferroglia:2012ku} is that these two extractions do not coincide!

Using our new result of the NNLO massive soft function, it is for the first time that one can directly validate the factorization formula (\ref{eq:sfac}) at the NNLO, and extract the soft-collinear fragmentation function $\tilde{s}_D$ at this order. It is then possible to resolve the conflict between the two results of $C_D$. In order to do these, we need to take the limit $m_t \to 0$ or $\beta \to 1$, and carefully extract the logarithms of $m_t$. Such logarithms arise from GPLs which are singular in the limit $\beta \to 1$. In order to extract such singularities, we employ properties of GPLs to convert their argument to $\cos\theta$, and put all $\beta$ dependences into the weights. We have used the program package \texttt{HyperInt} \cite{Panzer:2014caa} to accomplish this. After this conversion, the singular terms become powers of $\ln(1-\beta) \approx \ln(2m_t^2/M^2)$. Finally, we insert the results and the massless soft function from \cite{Ferroglia:2012uy} into Eq.~(\ref{eq:sfac}), and we find the NNLO coefficient of $\tilde{s}_D$ to be
\begin{align}
\tilde{s}_D^{(2)}(L_D,\mu) &= \frac{8}{9} L_D^4 + \left( \frac{76}{9} - \frac{8}{27} N_l \right) L_D^3 + \left( -\frac{104}{9} + \frac{76\pi^2}{27} + \frac{16}{27} N_l \right) L_D^2 \nonumber
\\
& + \left( \frac{440}{27} + \frac{416\pi^2}{27} - 72\zeta_3 + \frac{16}{81} N_l - \frac{16\pi^2}{27} N_l \right) L_D \nonumber
\\
& - \frac{1304}{81} - \frac{89\pi^2}{9} + \frac{1213\pi^4}{405} - \frac{1132\zeta_3}{9} + \left( -\frac{16}{243} + \frac{14\pi^2}{27} + \frac{88\zeta_3}{27} \right) N_l \, ,
\end{align}
The above formula differs from that in \cite{Gardi:2005yi, Neubert:2007je} by a constant term $4\pi^2C_AC_F$, which is essentially the inconsistency discussed in the Appendix~A of \cite{Ferroglia:2012ku}. Accordingly, the NNLO coefficient of the function $C_D$ is given by
\begin{align}
C_D^{(2)}(L_m,\mu) &= \frac{8}{9} L_m^4 + \left( \frac{20}{3} - \frac{8}{27} N_l \right) L_m^3 + \left( \frac{406}{9} - \frac{28\pi^2}{27} - \frac{52}{27} N_l \right) L_m^2 \nonumber
\\
&+ \left( \frac{2594}{27} + \frac{248\pi^2}{27} - \frac{232\zeta_3}{3} - \frac{308}{81} N_l -\frac{16\pi^2}{27} N_l \right) L_m \nonumber
\\
&+ \frac{21553}{162} + \frac{59\pi^2}{3} - \frac{749\pi^4}{405} + \frac{260\zeta_3}{9} + \frac{16\pi^2}{9} \ln 2 - \left( \frac{1541}{243} + \frac{74\pi^2}{81} + \frac{104\zeta_3}{27} \right) N_l \, ,
\end{align}
where $L_m \equiv \ln\big(\mu^2/m_t^2\big)$. The above expression of $C_D$ satisfies both Eq.~(\ref{eq:Dfac}) and the high-energy limit of the scattering amplitude. The inconsistency on the form of $C_D$ found in the Appendix~A of \cite{Ferroglia:2012ku} is thus resolved by our calculation. The remaining discrepancy then boils down to the relation between the $B$-meson shape function and the soft fragmentation function $\tilde{s}_D$. It would be interesting to directly compute $\tilde{s}_D^{(2)}$ from its operator definition in the future, in order to figure out its difference with the shape function.

\subsection{Numerical results}

The contribution of the soft function to the differential cross section is through the factorization formula in the soft limit \cite{Kidonakis:1996aq, Kidonakis:1997gm, Ahrens:2010zv}
\begin{align}
\frac{ d\tilde{\sigma} \big( N \big) }{dM \, d\cos\theta} \propto \sum_{ij=q\bar{q},gg} \mathrm{Tr} \! \left[ \bm{H}_{ij} \! \left( \ln\frac{M^2}{\mu^2},\beta,\cos\theta,\mu \right) \tilde{\bm{s}}_{ij} \! \left( \ln\frac{M^2}{\bar{N}^2\mu^2},\beta,\cos\theta,\mu \right) \right] + \mathcal{O}(1/N) \, ,
\end{align}
where $\bar{N} = Ne^{\gamma_E}$ with $N=M/\Lambda$ being the moment variable, and $\bm{H}_{ij}$ are the hard functions, whose matrix elements at the leading order are given by
\begin{align}
\bm{H}_{q\bar{q}}^{(0)} &=
\begin{pmatrix}
0 & 0
\\
0 & 2
\end{pmatrix}
\Bigg[ \frac{t_1^2 + u_1^2}{M^4} + \frac{2m_t^2}{M^2} \Bigg] \, , \nonumber
\\
\bm{H}_{gg}^{(0)} &=
\begin{pmatrix}
\frac{1}{N^2} & \frac{1}{N}\,\frac{t_1-u_1}{M^2} & \frac{1}{N}
\\
\frac{1}{N}\,\frac{t_1-u_1}{M^2} & \frac{(t_1-u_1)^2}{M^4} & \frac{t_1-u_1}{M^2}
\\
\frac{1}{N} & \frac{t_1-u_1}{M^2} & 1
\end{pmatrix}
\frac{M^4}{2t_1u_1} \Bigg[ \frac{t_1^2+u_1^2}{M^4} + \frac{4m_t^2}{M^2} - \frac{4m_t^4}{t_1u_1} \Bigg] \, .
\end{align}

In order to assess the numerical impact of the NNLO correction to the soft function, we define the following quantities:
\begin{align}
\mathcal{S}_{ij}^{(n)}(\beta,\mu/\mu_{\text{def}}) =  \int_{-1}^1 d\cos\theta \, \left( \frac{\alpha_s}{4\pi} \right)^n \mathrm{Tr} \! \left[ \bm{H}^{(0)}_{ij} \! \left( \beta,\cos\theta \right) \tilde{\bm{s}}^{(n)}_{ij} \! \left( \ln\frac{\Lambda^2}{\mu^2},\beta,\cos\theta \right) \right] ,
\end{align}
with $n=0$, 1, 2 denoting the LO, NLO and NNLO soft contributions, respectively. Here, $\mu_{\text{def}}$ is the default soft scale, and we exploit two kind of choices for it: $\mu_{\text{def},1} = \Lambda$ (which corresponds to the choice made in \cite{Pecjak:2016nee}) and $\mu_{\text{def},2} = \Lambda \sqrt{1-\beta^2\cos^2\theta}$ (which was found to be a better choice in \cite{Czakon:2018nun}). We then define the ratios
\begin{align}
R_{ij}^{\text{NLO}}(\beta,\mu/\mu_{\text{def}}) = \frac{\mathcal{S}_{ij}^{(1)}(\beta,\mu/\mu_{\text{def}})}{\mathcal{S}_{ij}^{(0)}(\beta,\mu/\mu_{\text{def}})} \, , \quad R_{ij}^{\text{NNLO}}(\beta,\mu/\mu_{\text{def}}) = \frac{\mathcal{S}_{ij}^{(2)}(\beta,\mu/\mu_{\text{def}})}{\mathcal{S}_{ij}^{(0)}(\beta,\mu/\mu_{\text{def}})} \, ,
\end{align}
to quantify the relative size of the NLO and NNLO soft corrections with respect to the LO contribution. In reality, the strong coupling $\alpha_s$ should also be evaluated at the soft scale $\mu$, and hence depends on the moment variable $N$. Here for illustration purposes, we fix its value at $\alpha_s = 0.118$. This does not affect the qualitative behaviors of the soft corrections shown below.

\begin{figure}[t!]
\begin{center}
\includegraphics[width=0.4\textwidth]{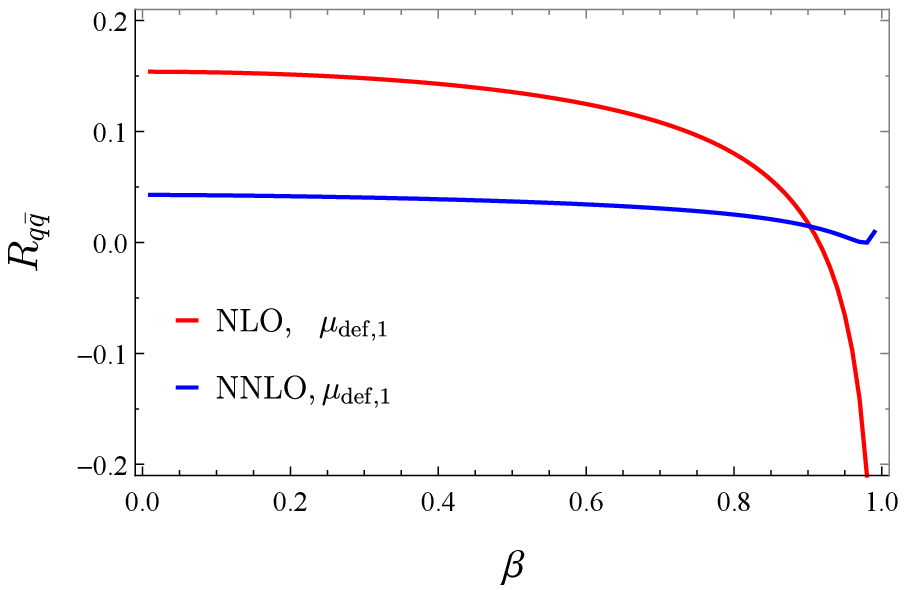}
\quad
\includegraphics[width=0.4\textwidth]{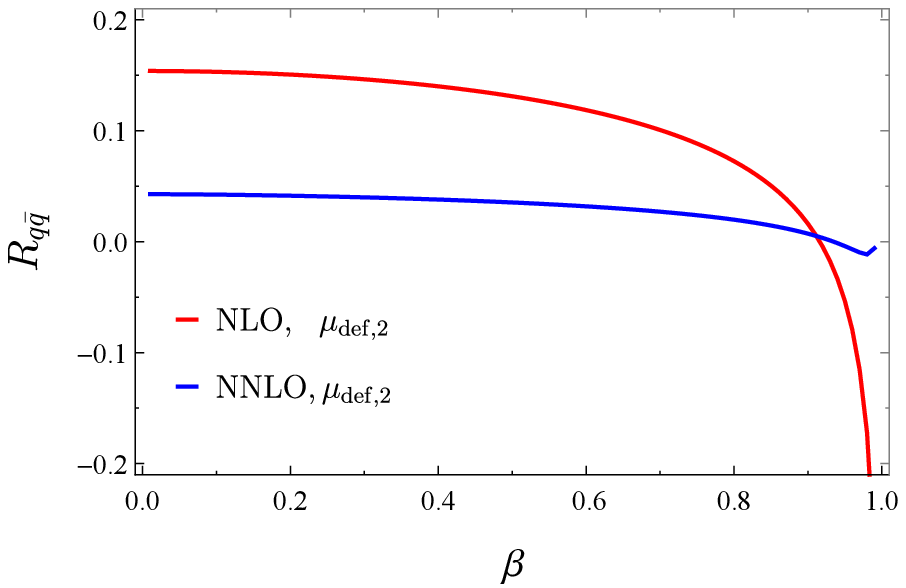}
\\[1ex]
\includegraphics[width=0.4\textwidth]{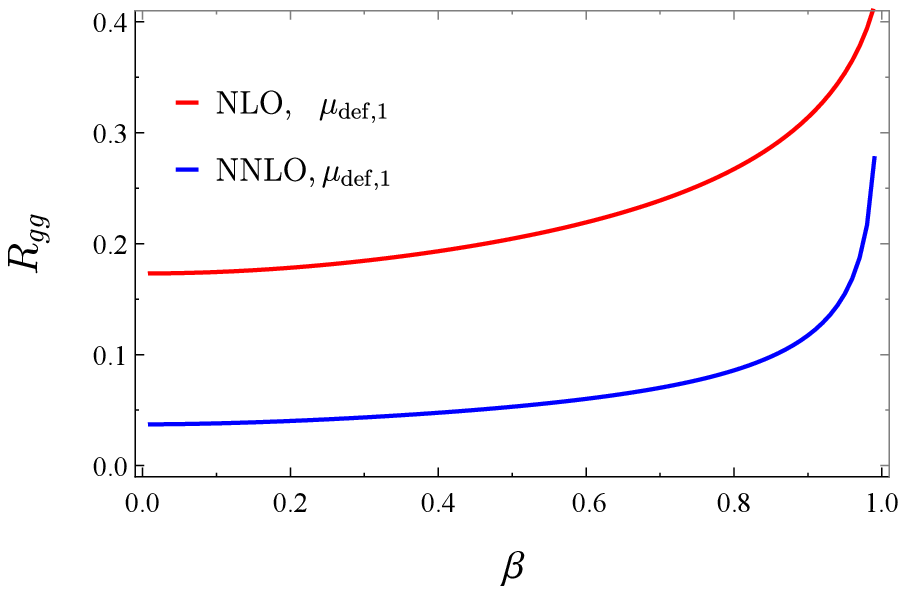}
\quad
\includegraphics[width=0.4\textwidth]{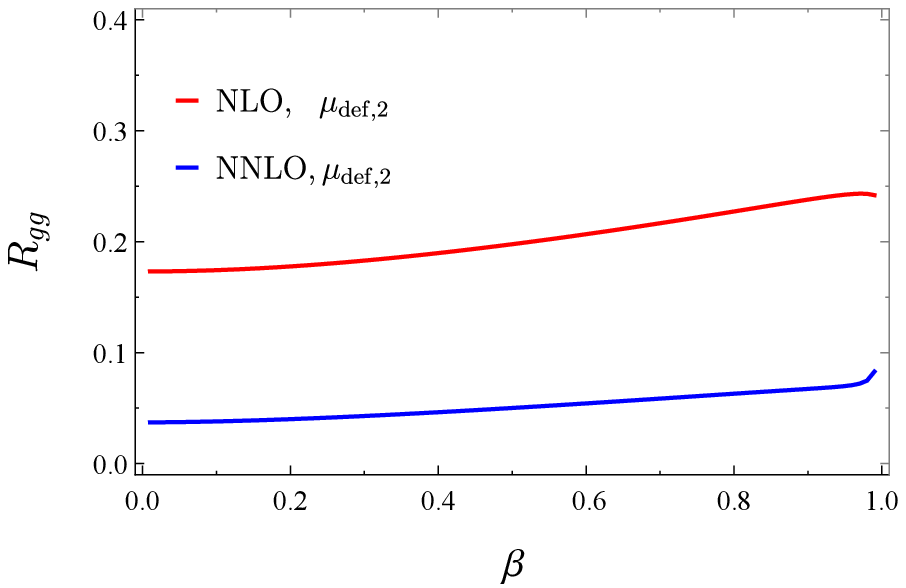}
\end{center}
\vspace{-2ex}
\caption{\label{fig:Sbeta}Relative soft corrections as a function of $\beta$ for two choices of the default scale: $\mu_{\text{def},1} = \Lambda$ (left) and $\mu_{\text{def},2} = \Lambda \sqrt{1-\beta^2\cos^2\theta}$ (right).}
\end{figure}

In Figure~\ref{fig:Sbeta}, we show the relative soft corrections as a function of $\beta$ for $\mu = \mu_{\text{def}}$, with the two choices of $\mu_{\text{def}}$. It can be seen that in the $q\bar{q}$ channel, the soft function is not very sensitive to the choice of the default soft scale, and the NNLO correction stays below 5\% in the whole range of $\beta$. On the other hand, the behavior of the $gg$ channel soft function is rather different. With the choice $\mu_{\text{def}} = \Lambda$, both the NLO and NNLO corrections become very large in the boosted limit $\beta \to 1$, where the NNLO correction can reach about 28\%. When changing to the choice $\mu_{\text{def}} = \Lambda \sqrt{1-\beta^2\cos^2\theta}$, the corrections are well under control in the boosted region, where the NNLO correction is only about 8\%. These findings are in coincidence with the discussions in \cite{Czakon:2018nun}, where the second choice was identified as the better choice, especially for boosted top quark pair production.

\begin{figure}[t!]
\begin{center}
\includegraphics[width=0.4\textwidth]{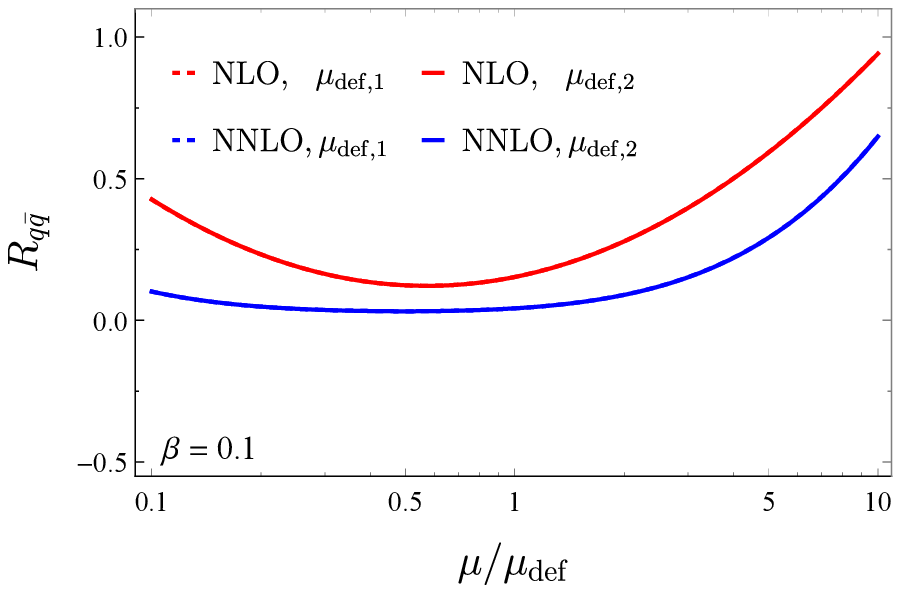}
\quad
\includegraphics[width=0.4\textwidth]{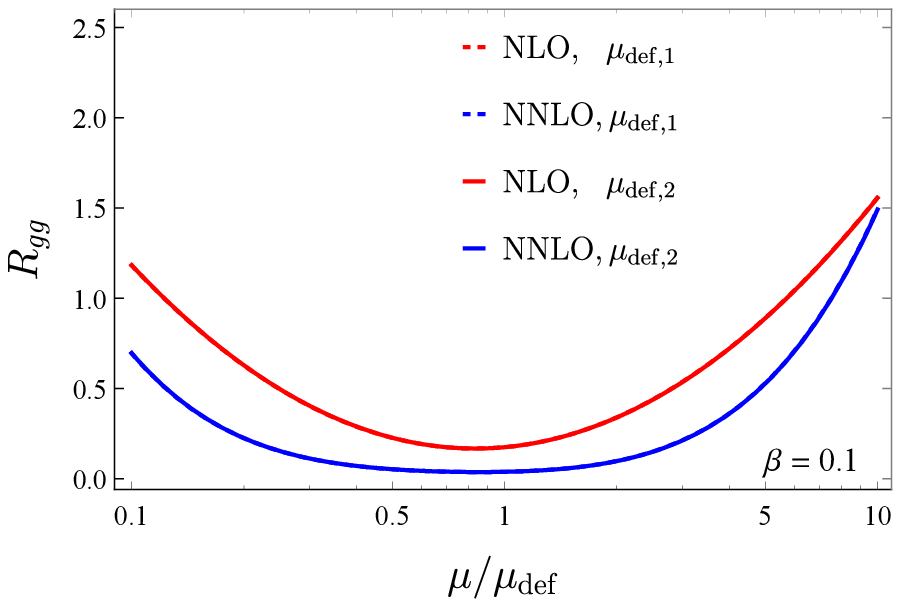}
\\[0.5ex]
\includegraphics[width=0.4\textwidth]{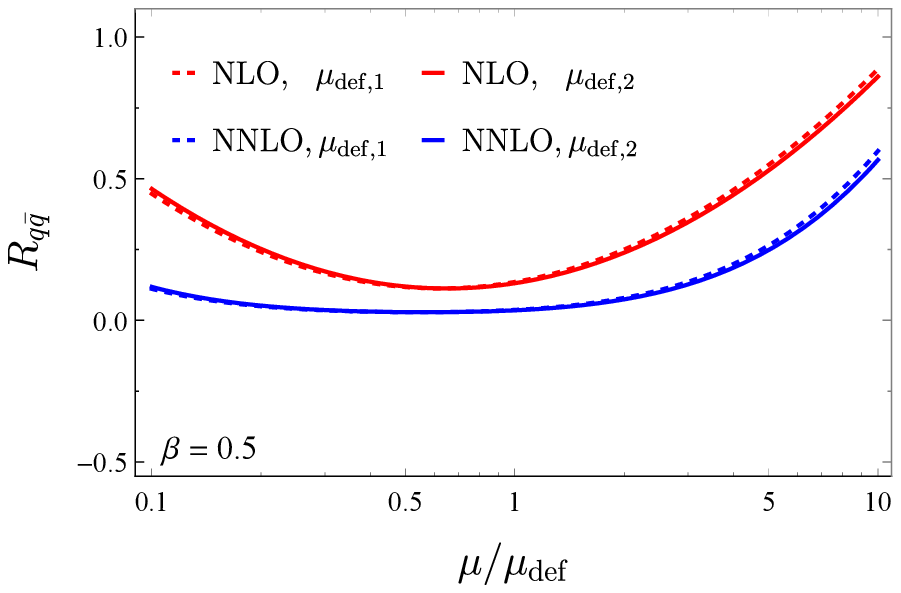}
\quad
\includegraphics[width=0.4\textwidth]{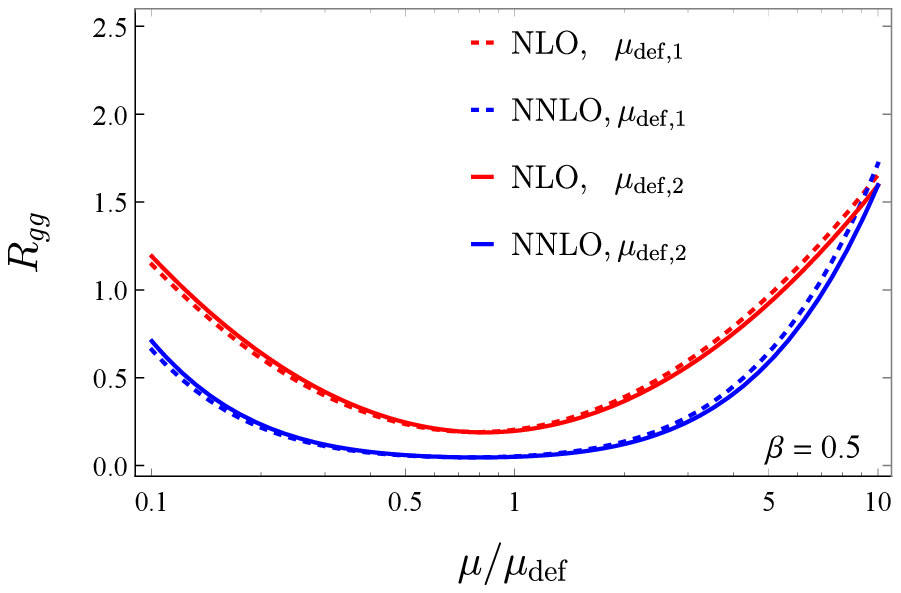}
\\[0.5ex]
\includegraphics[width=0.4\textwidth]{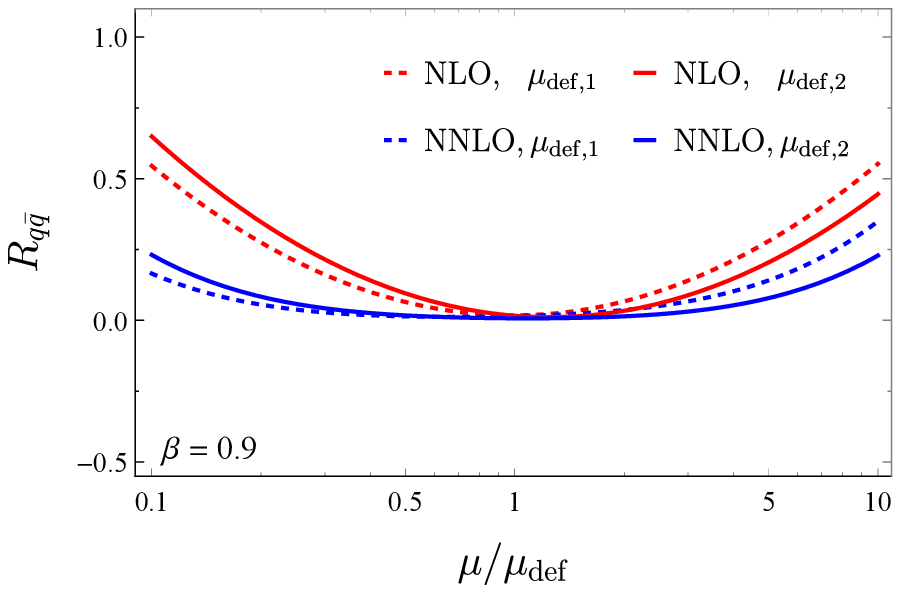}
\quad
\includegraphics[width=0.4\textwidth]{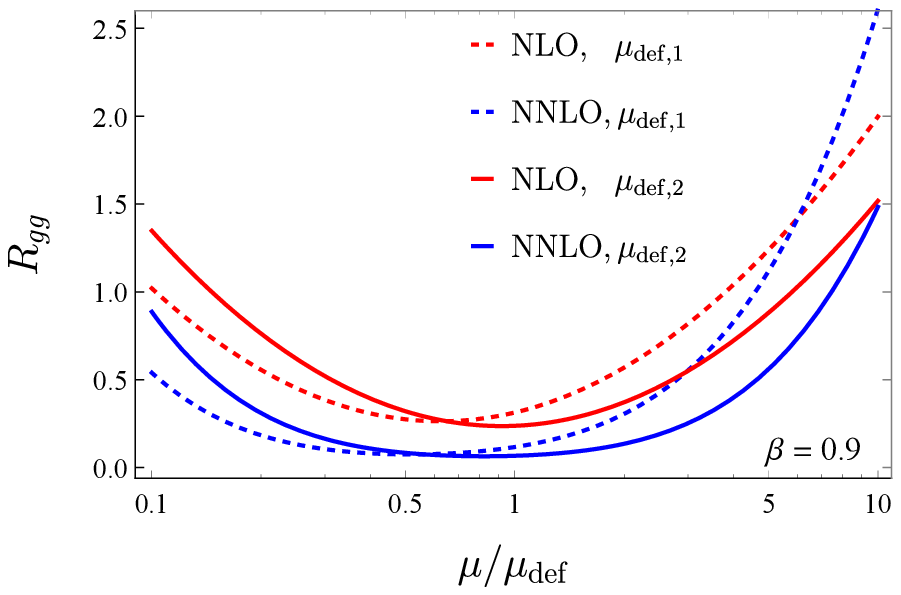}
\\[0.5ex]
\includegraphics[width=0.4\textwidth]{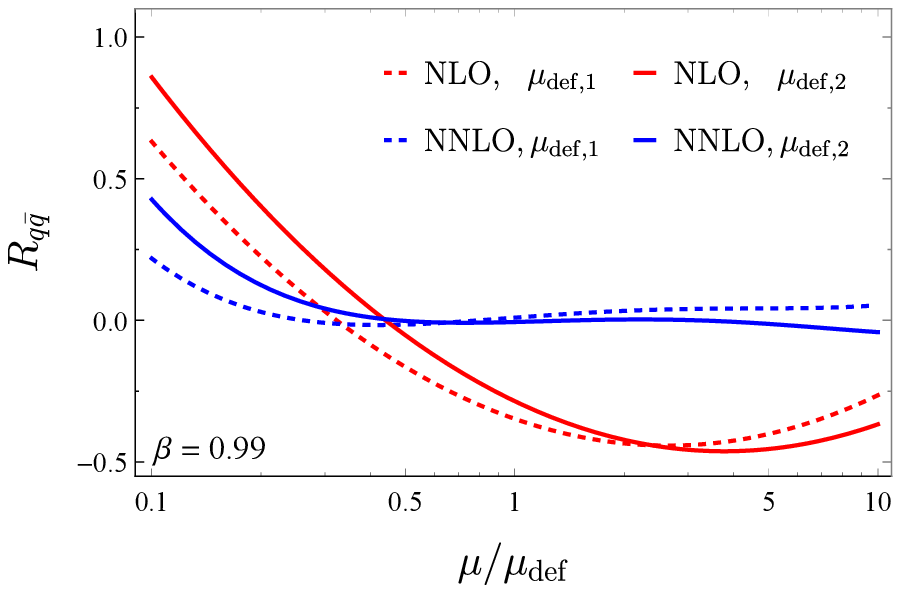}
\quad
\includegraphics[width=0.4\textwidth]{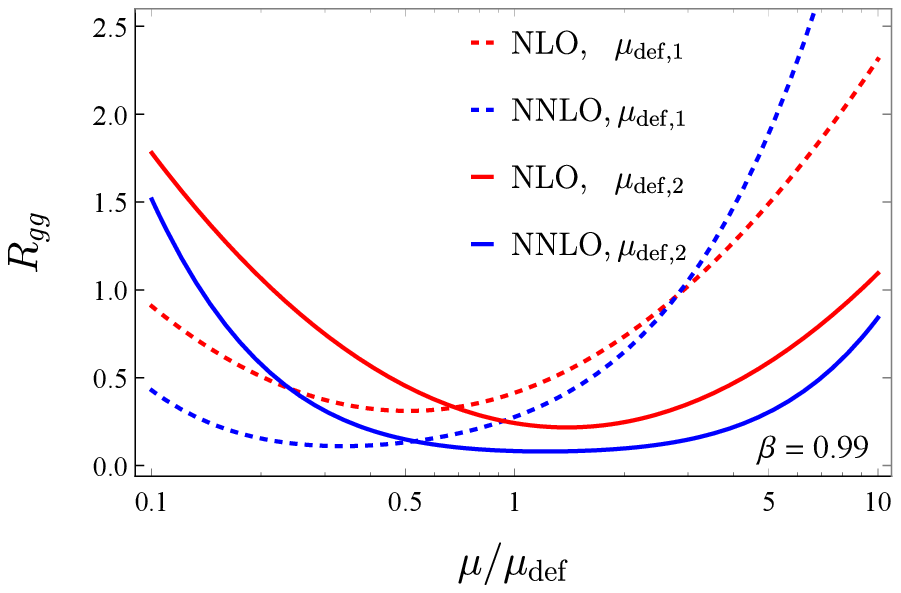}
\end{center}
\vspace{-2ex}
\caption{\label{fig:Smu}Relative soft corrections as a function of $\mu/\mu_{\text{def}}$ for two choices of the default scale: $\mu_{\text{def},1} = \Lambda$ (dashed curves) and $\mu_{\text{def},2} = \Lambda \sqrt{1-\beta^2\cos^2\theta}$ (solid curves).}
\end{figure}

In Figure~\ref{fig:Smu}, we show the relative soft corrections as a function of $\mu/\mu_{\text{def}}$ for 4 values of $\beta$: $\beta = 0.1$ (the threshold region), $\beta = 0.5$ (the dominant region for the total cross section), $\beta = 0.9$ (the boosted region) and $\beta = 0.99$ (the ultra-boosted region). For $\beta = 0.1$, the two choices of the default soft scale are almost indistinguishable as the solid curves and the dashed curves overlap with each other in the first row of Figure~\ref{fig:Smu}. When $\beta$ gradually becomes larger, the two choices start to differ. We observe that the second option is a good choice in the whole range of $\beta$, in the sense that the soft corrections remain small when the soft scale is varied around the default value. This is particularly true for the $gg$ channel, where the $t$- and $u$-channel propagators in the tree-level hard function push the average value of $\cos^2\theta$ towards unity when the top quarks are highly boosted. In that case the effective soft scale is much smaller than $\Lambda$, and is better modeled by the $\theta$-dependent function $\mu_{\text{def},2} = \Lambda \sqrt{1-\beta^2\cos^2\theta}$.  While the above findings have been advocated in \cite{Czakon:2018nun} from studying the massless soft function, our analysis using the massive soft function provides more comprehensive information on the behavior of the soft corrections in the whole range of phase space.

\section{Conclusion and outlook}
\label{sec:con}

In this paper we have calculated the threshold soft function for top quark pair production at the NNLO. We used integration-by-parts identities to reduce the double-real contributions to 23 master integrals, and the virtual-real contributions to 36 master integrals. We then employed the method of differential equations to solve for the master integrals to arbitrary orders in the dimensional regulator $\epsilon$. Our final results are fully analytic, and can be entirely written in terms of GPLs, which makes it efficient for numerical evaluation. Our result represents the first ever NNLO soft function for processes involving a non-trivial color structure and two massive partons with full velocity dependence.

Due to the complicated color structure, the renormalized soft function calculated in this paper is a $2\times 2$ ($3\times 3$) matrix for $q\bar{q} \to t\bar{t}$ ($gg \to t \bar{t}$) in color space. The scale-dependence of the soft function is in full agreement with the two-loop anomalous dimension matrix calculated in \cite{Ferroglia:2009ep, Ferroglia:2009ii, Chien:2011wz} from virtual amplitudes, as expected from the cancellation of infrared singularities. However, the set up of the calculation for the soft function is different from the calculation for virtual amplitudes, and therefore represents an independent confirmation of previous results on the two-loop anomalous dimensions. We find that the previously calculated three-parton correlation function $f_2$ in the anomalous dimension comes entirely from virtual-real diagrams involving the two massive Wilson lines. This suggests a Coulomb/Glauber origin of the three-parton correlation. It would be interesting to investigate about this in the future.

Our result contains full velocity dependence of the massive partons, which generalizes previous results in more restricted kinematic configurations to fully generic configurations. We have checked that in the limit $\beta \to 0$, our result reproduces the corresponding results for color singlet/octet production. In the high energy limit $\beta \to 1$, our result exhibits the expected factorization property of mass logarithms, which leads to a consistent extraction of the soft fragmentation function. We also find full agreement with the NNLO massless soft function in \cite{Ferroglia:2012uy}, up to the three-parton virtual-real contributions not calculated in that paper.

Our result is an important ingredient in the resummation of threshold logarithms for top quark pair production beyond the NNLL accuracy. It is interesting to study its phenomenological implications for the LHC experiments and future high energy colliders, following the framework of \cite{Pecjak:2016nee, Czakon:2018nun}. For this purpose, we have studied the numerical impact of the NNLO corrections to the soft function. We find that using a $\theta$-dependent default choice for the soft scale $\mu_{\text{def}} = \Lambda \sqrt{1 - \beta^2 \cos^2\theta}$ makes the perturbative behavior well under-control in the whole phase space, from production at rest to highly-boosted production. This is in accordance with the findings of \cite{Czakon:2018nun}. To achieve the next-to-next-to-next-to-leading logarithmic accuracy for resummation, one also needs the two-loop hard function and the three-loop soft anomalous dimension matrix. The two-loop hard function can be extracted from the virtual amplitudes calculated in \cite{Baernreuther:2013caa}. Concerning the three-loop soft anomalous dimensions for multi-leg scattering processes, the result in the purely massless case has been obtained in \cite{Almelid:2015jia}, but the result in the massive case is still lacking and deserves future investigations.

Finally, while our formalism is generic, in practical calculation we have used the fact that the top quark and the anti-top quark are back-to-back in the partonic center-of-mass frame. It would be interesting to consider the more general case, where the $t\bar{t}$ pair is recoiled by additional particles. This is relevant to the production processes of top quark pairs associated with a Higgs boson or an electroweak gauge boson. At the NLO, the general result of \cite{Ahrens:2010zv} can be utilized, and has been successfully applied to $t\bar{t}H$ production in \cite{Broggio:2015lya, Broggio:2016lfj}, to $t\bar{t}W$ production in \cite{Li:2014ula, Broggio:2016zgg}, and to $t\bar{t}Z$ production in \cite{Broggio:2017kzi}. It remains open whether a similar general result can be derived at the NNLO, which we leave for future work.

\section*{Acknowledgments}

This work was supported in part by the National Natural Science Foundation of China under Grant No. 11575004 and 11635001.

\end{document}